\documentclass[useAMS,usenatbib]{mn2e}
\usepackage{epsfig}
\usepackage{amsmath}

\newcommand{\be}{\begin{equation}}
\newcommand{\beq}{\begin{equation}}
\newcommand{\ba}{\begin{eqnarray}}
\newcommand{\ee}{\end{equation}}
\newcommand{\eeq}{\end{equation}}
\newcommand{\ea}{\end{eqnarray}}

\def\lsim{~\rlap{$<$}{\lower 1.0ex\hbox{$\sim$}}}

\def\gsim{~\rlap{$>$}{\lower 1.0ex\hbox{$\sim$}}}

\title[Reionization in Galaxy Formation Models]{The Structure of Reionization in Hierarchical Galaxy Formation Models}
\author[Han-Seek Kim et al.]
       {Han-Seek~Kim$^{1}$\thanks{hansikk@unimelb.edu.au}, J. Stuart B.~Wyithe$^{1,3}$, Sudhir Raskutti$^{1}$, C. G.~Lacey$^{2}$ \newauthor and J. C.~Helly$^{2}$\\
       $^1$School of Physics, The University of Melbourne, Parkville, VIC 3010, Australia\\
      $^2$Institute for Computational Cosmology, Department of Physics, University of Durham, South Road, Durham DH1 3LE, UK\\
  	$^3$ARC Centre of Excellence for All-sky Astrophysics (CAASTRO)}
\date{}

\pagerange{\pageref{firstpage}--\pageref{lastpage}}
\pubyear{2011} 

\begin{document}

\maketitle

\label{firstpage}

\begin{abstract}

\noindent Understanding the epoch of reionization and the properties of the first galaxies represents an important goal for modern cosmology. The structure of reionization, and hence the observed power spectrum of redshifted 21cm fluctuations are known to be sensitive to the astrophysical properties of the galaxies that drove reionization. Thus, detailed measurements of the 21cm power spectrum and its evolution could lead to measurements of the properties of early galaxies that are otherwise inaccessible. In this paper, we make predictions for the ionised structure during reionization and the 21cm power spectrum based on detailed models of galaxy formation. We combine the semi-analytic GALFORM model implemented within the Millennium-II dark matter simulation, with a semi-numerical scheme to describe the resulting ionization structure. Semi-analytic models based on the Millennium-II simulation follow the properties of galaxies within halos of mass greater than $\sim$ 1.4$\times$$10^{8}$M$_\odot$ at $z>6$, corresponding to the faint sources thought to dominate reionization. Using these models we show that the details of SNe and radiative feedback affect the structure and distribution of ionised regions, and hence the slope and amplitude of the 21 cm power spectrum. These results indicate that forthcoming measurements of the 21cm power-spectrum could be used to uncover details of early galaxy formation. We find that the strength of SNe feedback is the dominant effect governing the evolution of structure during reionization. In particular we show SNe feedback to be more important than radiative feedback, the presence of which we find does not influence either the total stellar mass or overall ionising photon budget. Thus, if SNe feedback is effective at suppressing star formation in high redshift galaxies, we find that photoionization feedback does not lead to self-regulation of the reionization process as has been thought.

\end{abstract}


\section{Introduction}
 
In anticipation of forthcoming 21~cm observations of the epoch of reionization, a great deal of
theoretical attention has focused on the prospects of measuring the 21~cm power spectrum. 
To this end, significant progress has been made in modeling the effect of galaxies on
the reionization of the IGM. In large modern simulations, the most common approach
 is to begin with an N-body code to generate a distribution of
halos \citep[e.g.][]{ciardi2003,sokasian2003,iliev2007,zahn2007,trac2007,shin2008,Il08,trac2008}. A simple prescription is then used to relate halo mass to ionizing luminosity. Following this step, radiative transfer methods (most commonly ray-tracing algorithms) are employed to model the generation of ionized structure on large scales. The radiative transfer is normally run with lower resolution than the N-body code for computational
efficiency.

These simulations describe the generic features of reionization
\citep[e.g.][]{iliev2007,zahn2007,mcquinn2007,shin2008,lee2008,croft2008}, confirming expectations from analytic models \citep[e.g.][]{furl2004a,furl2004b,WM07,barkana2008} that large-scale,
over-dense regions near sources are generally reionized first, and that massive galaxies tend to be surrounded by clustered sources that increase the size of HII regions. In addition, the simulations describe the structure of the HII regions, showing that they are generally aspherical (even where the sources are assumed to emit isotropically). The growth of HII regions during reionization may also be influenced by radiative feedback in
the form of suppression of galaxy formation below the cosmological Jeans mass within a heated IGM
\citep{dijkstra2004}, although the importance of this effect remains controversial \citep{mesinger2008}. Suppression of low mass galaxy-formation delays and extends the reionization process, which though started by low mass galaxies, must then be completed by relatively massive galaxies~\citep{iliev2007}. 

An important outcome from the large cosmological volumes attained by 
modern numerical simulations has been the prediction of 21~cm
signals that will be observable using forthcoming low frequency arrays
\cite[e.g.][]{mellema2006,lidz2008}. The most generic features of 21~cm power spectrum modeling 
were elucidated by \citet{lidz2008}, who show examples of its evolution. On scales of $k\sim0.1$Mpc$^{-1}$ the amplitude and the slope of the power spectrum vary in a non-monotonic way relative to the expected shape in the absence of ionization structure. Thus, measurement of the 21cm power spectrum will provide the first clues regarding the clustering of ionizing sources during reionization. In particular, \citet{lidz2008} illustrate that
the slope and amplitude of the 21~cm power spectrum vary considerably
among different models at a given ionization fraction. However
they also find that the behavior with ionization fraction across the
different models is relatively generic. In particular, the amplitude of the 21~cm
power spectrum reaches a maximum close to the epoch when $\sim 50\%$
of the volume of the IGM is ionized, while its slope is found to flatten with
increasing ionization fraction. \citet{lidz2008} argue that 
first generation low frequency radio telescopes like the Murchison Widefield Array\footnote{http://www.haystack.mit.edu/ast/arrays/mwa/} and the Low Frequency Array\footnote{http://www.lofar.org/} will have sufficient sensitivity to measure the
redshift evolution in the slope and amplitude of the 21~cm power
spectrum. 

One of the main limitations in modelling of reionization is the physics of the ionizing sources. Most studies have used very simple prescriptions to assign ionizing luminosities to dark matter halos. It has been shown that it is then possible to constrain the parameters for these simple prescriptions. However an important open question is the degree to which the important astrophysics governing formation and evolution of high redshift galaxies is accessible via observations of the 21cm power spectrum. A few studies have previously addressed the issue of realistic modelling of high redshift galaxies. For example, \citet[][]{Theuns2011} \citep[see also ][]{Benson2006,Lacey2011} used the semi-analytical galaxy formation code, GALFORM \citep{Cole2000,Baugh2005,Bower2006}, based on Monte-Carlo merger trees to evaluate the ionizing photon budget, finding that although galaxies should produce sufficient ionizing photons to complete reionization, most of the galaxies responsible would be below the detection threshold of current surveys. Furthermore \citet[][]{Theuns2011} and \citet[][]{Benson2006} have studied the effect of SN feedback on the global ionizing photon budget and global ionization. However these studies were restricted to the global evolution, and do not address the ionization structure.

In this paper we combine detailed models of high redshift galaxy formation using GALFORM with calculations of the spatial dependence of reionization, and predict the resulting redshifted 21cm power spectrum. We begin in \S~\ref{modell} and \S~\ref{scheme} by describing the implementation of GALFORM, and our method for modelling the ionization structure. Then, in \S~\ref{Maps} we present ionisation maps for different galaxy formation models, including the effect of SNe and radiative feedback. We discuss the ionising photon budget as a function of halo circular velocity in \S~\ref{budget}, and then present a discussion of the dependence of the 21cm power spectrum on the galaxy formation model in \S~\ref{PS}.  We finish with some conclusions in \S~\ref{Summary}.

\section{The model} \label{modell}
In this section we introduce the theoretical galaxy formation modelling used in our analysis. In \S~\ref{GFM}, we briefly review GALFORM. We then describe the implementations of SNe, AGN and photoionization feedback processes in \S~\ref{Feedbacks}. 

\subsection{The GALFORM galaxy formation model}\label{GFM}

The formation and evolution of galaxy properties are computed within the $\Lambda$CDM structure formation framework using the semi-analytical model GALFORM. GALFORM includes a range of processes that are thought to be important for galaxy formation, including: (1) The gravitationally driven assembly of dark matter haloes; (2) The density and angular
momentum profiles of dark matter and hot gas in haloes; (3) The radiative cooling of gas and
its collapse to form centrifugally supported disks; (4) Star formation in disks; (5)
Feedback processes, resulting from the injection of energy from Supernovae (SNe)
and AGN heating; (6) Chemical enrichment of the interstellar medium (ISM) and
hot halo gas which affect the gas cooling rate and the properties of the stellar populations in a galaxy; (7) The dynamical friction on orbits of satellite galaxies within a dark matter
halo and their possible merger with the central galaxy; (8) The formation of galactic
spheroids; (9) The spectrophotometric evolution of stellar populations; (10) The
effect of dust extinction on galaxy luminosities and colours, and its dependence on
the inclination of a galaxy; (11) The generation of emission lines from interstellar
gas ionized by young hot stars. 

A comprehensive overview of GALFORM can be found in \citet[][]{Cole2000}, with an updated discussion in the review by \citet[][]{Baugh2006}. In this paper, we implement GALFORM within the Millennium-II cosmological N-body simulation \citep[]{MII2009}. However rather than using the halo merger trees presented in \cite{MII2009}, we base our study on the halo merger trees described in the study of \citet[][]{Merson2012} { which are better suited for the purposes of semi-analytic modelling}. The simulation has a cosmology including fractional mass and dark energy densities with values of $\Omega_{\rm m}=0.25$ , $\Omega_{\rm b}=0.045$ and $\Omega_{\Lambda}$=0.75, a dimensionless Hubble constant of $h$=0.73, and a power spectrum normalisation of $\sigma_{8}$=0.9. The particle mass of the simulation is 6.89$\times$10$^{6}$h$^{-1}{\rm M_{\odot}}$ and we detect haloes down to 20 particles in the simulation box of side length $L=100h^{-1}$Mpc. 
%The resolution of the simulation is fixed at a halo mass of $\sim$10$^{8}$$h^{-1}{\rm M_{\odot}}$

\subsection{Feedback processes}
\label{Feedbacks}

Feedback processes during galaxy formation are very important contributors to the shape of luminosity functions predicted by GALFORM \citep[][]{Cole2000,Benson2002,Baugh2005,Bower2006,Kim2011}. Three main feedback processes are implemented in GALFORM, which we discuss in turn below. 

\subsubsection{SNe feedback}\label{Sne}
SNe feedback on galaxy formation is implemented within GALFORM through the reheating and ejection of cold gas from galaxies via the equation  
\begin{equation}
\dot{M}_{\rm eject}=\beta\psi,
\end{equation}
where $\psi$ is the instantaneous star formation rate. Here $\beta$ is the efficiency of the feedback process, parameterized as 
\begin{equation}\label{snbeta}
\beta=(V_{\rm disk}/V_{\rm hot})^{-\alpha_{\rm hot}},
\end{equation}
{ where $V_{\rm hot}$ has unit of $\rm{kms^{-1}}$, $\alpha_{\rm hot}$ is a dimensionless adjustable parameter which controls the strength of SN feedback \citep[see ][]{Cole2000} and $V_{\rm disk}$ is the circular velocity of the galactic disk at the half-mass radius.} SNe feedback suppresses the formation of galaxies within small dark matter halos, and is required to reproduce the faint end of the observed galaxy luminosity function \citep[e.g.][]{Norberg2002,Blanton2001}. \citet[][hereafter the Bow06 model]{Bower2006} adopted values of $V_{\rm hot}$=485km/s and $\alpha_{\rm hot}$=3.2. 

\subsubsection{AGN feedback}
\label{AGN}

To reproduce the low number density of bright galaxies and the steep slope of the bright end of the galaxy luminosity function an additional feedback process is needed that operates in the high mass regime. For this reason, Bow06 included AGN feedback which suppresses the cooling flows in massive haloes. The physical motivation for this lies in the fact that energy is known to be released from accretion of matter onto central supermassive black holes. Bow06 modelled AGN feedback assuming halos to be in quasi-hydrostatic equilibrium in cases where the cooling time at the cooling radius, $t_{\rm cool}(r_{\rm cool})$, exceeds a multiple of the free-fall time at the cooling radius, $t_{\rm ff}(r_{\rm cool})$, i.e.
\begin{equation}
t_{\rm cool}(r_{\rm cool})>{1 \over \alpha_{\rm cool}} t_{\rm ff}(r_{\rm cool}),
\end{equation}
where $\alpha_{\rm cool}$ is an adjustable parameter whose value controls the strength of AGN feedback. The value of $\alpha_{\rm cool}$ in the Bow06 model is 0.58. An alternative approach was taken by \citet[][hereafter Bau05]{Baugh2005} who implemented superwind feedback from star formation with a top-heavy stellar initial mass function in order to understand the high luminosity end of the galaxy luminosity function.

\subsubsection{Photoionization feedback}
\label{ionization}

In the presence of a strong ionizing background, star formation in small galaxies is quenched owing to several physical processes, including the suppression of cooling by photo-heating \citep[][]{efstathiou1992}, the higher IGM gas pressure \citep[][]{gnedin2000}, and photo-heating \citep[][]{Hoeft2006,Okamoto2008}. As a result the star formation rate density is suppressed within HII regions during reionization \citep[see ][]{Crain2009}, which may result in self-regulation of the reionization process \citep{iliev2007}. Based on \citet[][]{Benson2002}, GALFORM includes a prescription for suppressing the cooling of 
halo gas onto the galaxy when { the IGM becomes globally ionized. In the standard implementation this is assumed to occur at a particular redshift 
$z_{cut}$. It is assumed that the suppression of cooling 
occurs when the host halo's circular velocity lies below a threshold 
value, $V_{\rm cut}$, at redshift $z<z_{\rm cut}$. Rather than a constant $z_{\rm cut}$, in this paper we apply suppression when the cell in which the galaxy resides is fully ionized (see  \S~\ref{SM}).} The adopted value is $V_{\rm cut}=30\,$km/s for each of the models \citep[][]{Lacey2011,Lagos2012}. We note that the value of $V_{\rm cut}=30\,$km/s is used for the Bow06 model in this paper, rather than the original value $V_{\rm cut}=50\,$km/s used in \citep[]{Bower2006}.

%%%%%%%%%%%%%%

\subsubsection{Key differences of variant models}
\label{variantmodels}

In this paper, we use variants of the GALFORM galaxy
formation model described in \S~\ref{GFM} based on two main
published implementations which we refer to as  Bow06 \citep{Bower2006}
and Lagos \citep{Lagos2012}. The Bow06 and Lagos models assume a
Kennicutt IMF, similar to that in the solar neighbourhood, for both
quiescent star formation and starbursts. The Lagos model is similar to the Bow06 model
but it has a new star formation law and has different photoionization
feedback parameters. We study three variants of the Bow06 model, in addition to the published version. First, the NOSN
model in which we remove SNe feedback from the Bow06 model by using a value of $V_{\rm hot}$=0km/s. Second, the Bow06(no suppression) model in which we remove photo-ionization by setting $V_{\rm cut}$=0km/s, and third the NOSN(no suppression) model in which we remove both SNe and radiative feedback (i.e. $V_{\rm hot}=V_{\rm cut}$=0km/s). In Table~\ref{Parameters}, we summarize the values of selected parameters for the different models used in this paper.

\begin{table*}
\caption{
The values of selected parameters which are different in the models. The columns are as follows: (1) the name of the model, (2) the value of the photoionization parameter $V_{\rm{cut}}$, (3) the SNe feedback parameter, $V_{\rm{hot}}$, 
(4) the IMF of brown dwarfs $\Upsilon$, and (5) comments giving model source or key differences from published models.}
\label{Parameters}
\begin{tabular}{lcccl}
\hline
\hline
 & $V_{\rm{cut}}$[kms$^{-1}]$&$V_{\rm{hot}}$[kms$^{-1}$]& $\Upsilon$ & Comments\\
\hline
Bow06 & 30  & 485  & 1 & Bower et~al. (2006), $V_{\rm{cut}}$ value change \\
Lagos& 30  &  485  & 1   & Lagos et~al. (2012)\\
Bow06(no suppression)& 0 & 485 & 1 &  Bower~et al. (2006), No radiative suppression\\
NOSN& 30 & 0 & 4       & Bower~et al. (2006), No SNe feedback \\
NOSN(no suppression)&0& 0 & 4 & Bower~et al. (2006)\\
 & & & & No SNe feedback and No radiative suppression\\
\hline
\end{tabular}
\end{table*}

\subsection{Modelling spatial dependence of radiative feedback in GALFORM}
\label{SM}

Photoionization feedback from reionization is normally modelled in semi-analytic models (including GALFORM) using a single value of $z_{\rm cut}$. However to investigate the effect of galaxy formation on ionisation structure during reionization we need to improve the photo-ionizing feedback using a spatially dependent value of $z_{\rm cut}$ that accounts for earlier suppression in regions of the IGM where HII regions first form.   Broadly, the process is to run GALFORM to a particular redshift snapshot, perform a calculation of ionisation structure, and then apply radiative suppression to subsequent galaxy formation inside HII regions with halo circular velocities below $V_{\rm cut}$. GALFORM then evolves the population of galaxies to the next snapshot, where the ionisation structure is recomputed. Details of the GALFORM specific implementation are provided in the Appendix.

\subsection{The High redshift galaxy luminosity function}

\begin{figure*}
\begin{center}
\includegraphics[width=7cm]{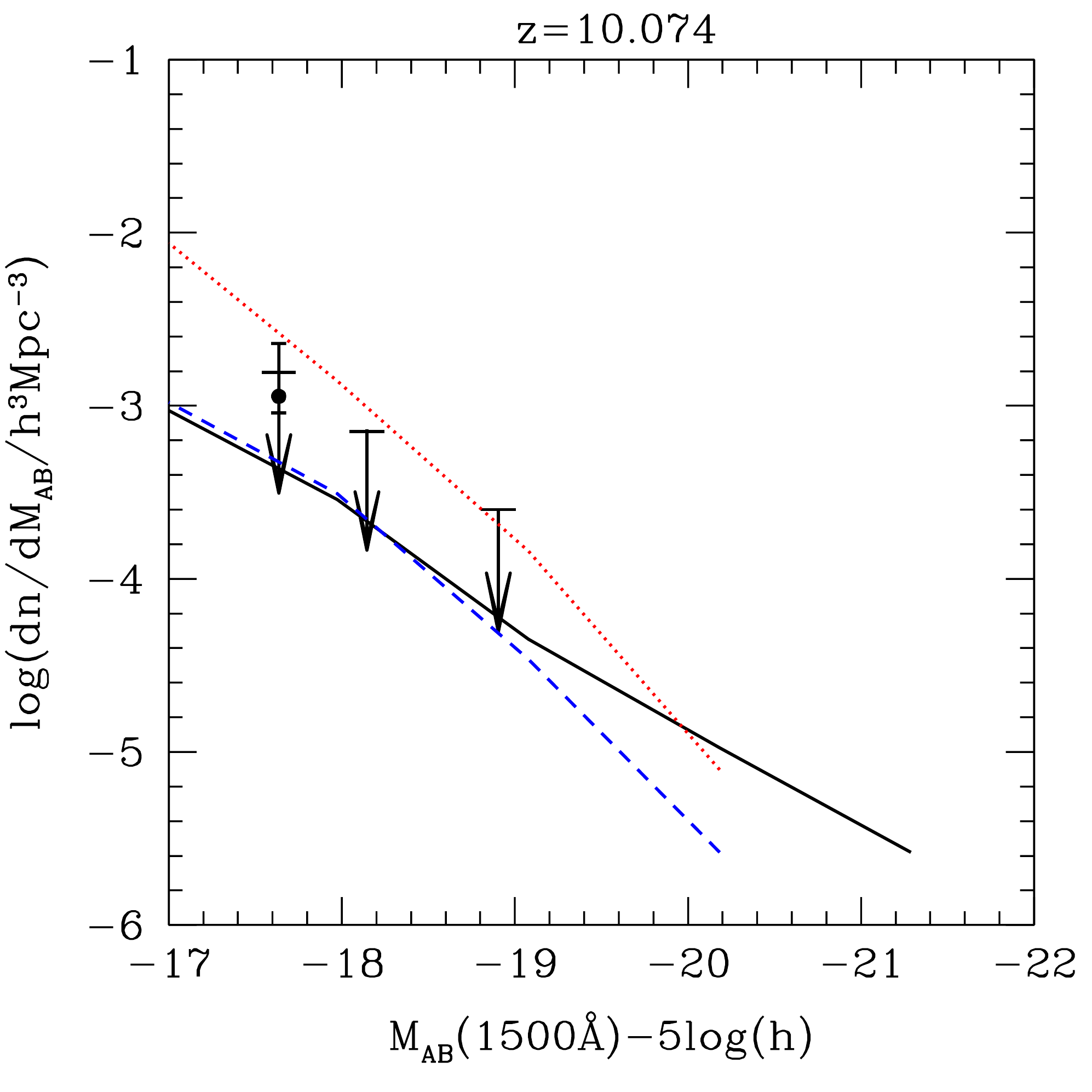}
\includegraphics[width=7cm]{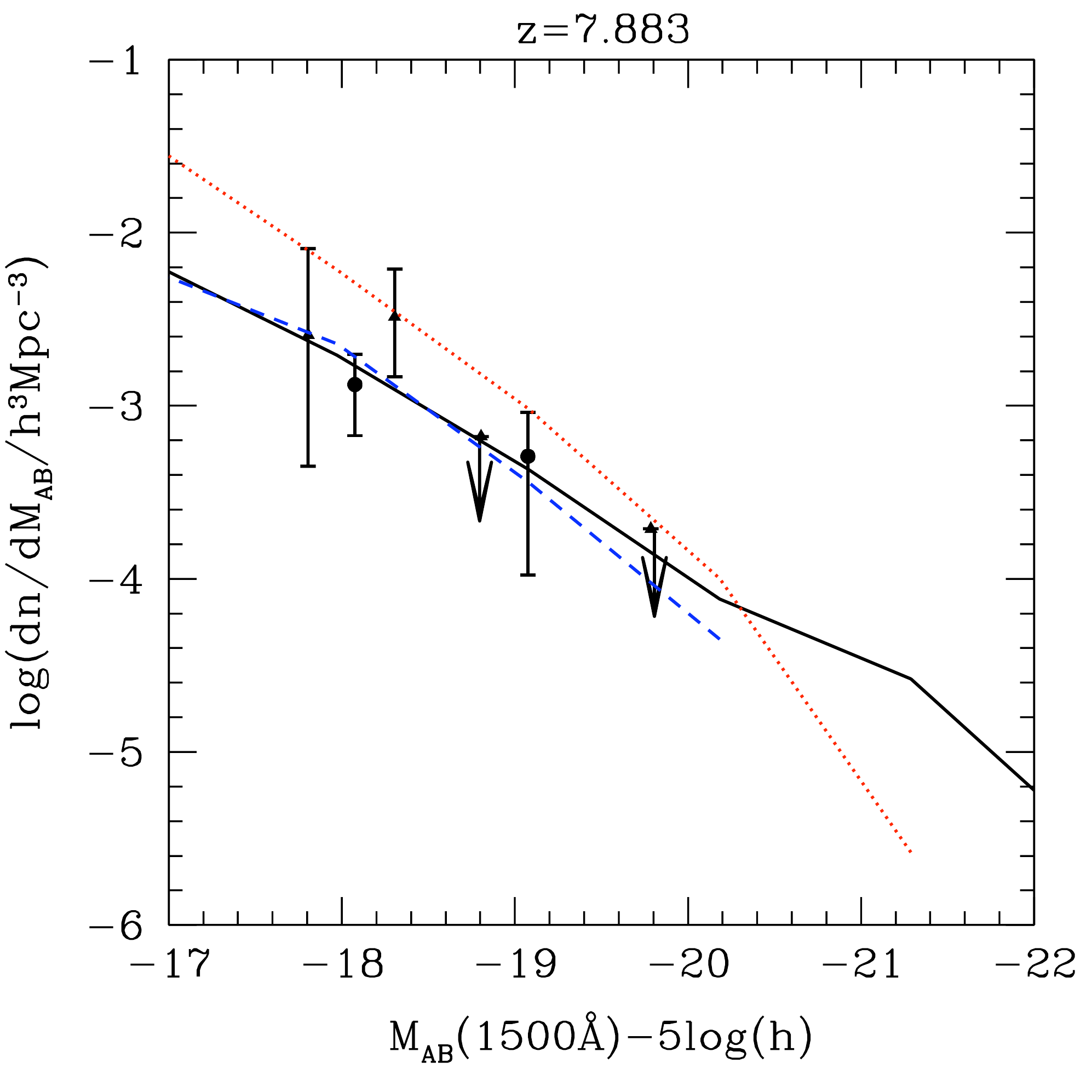}
\includegraphics[width=7cm]{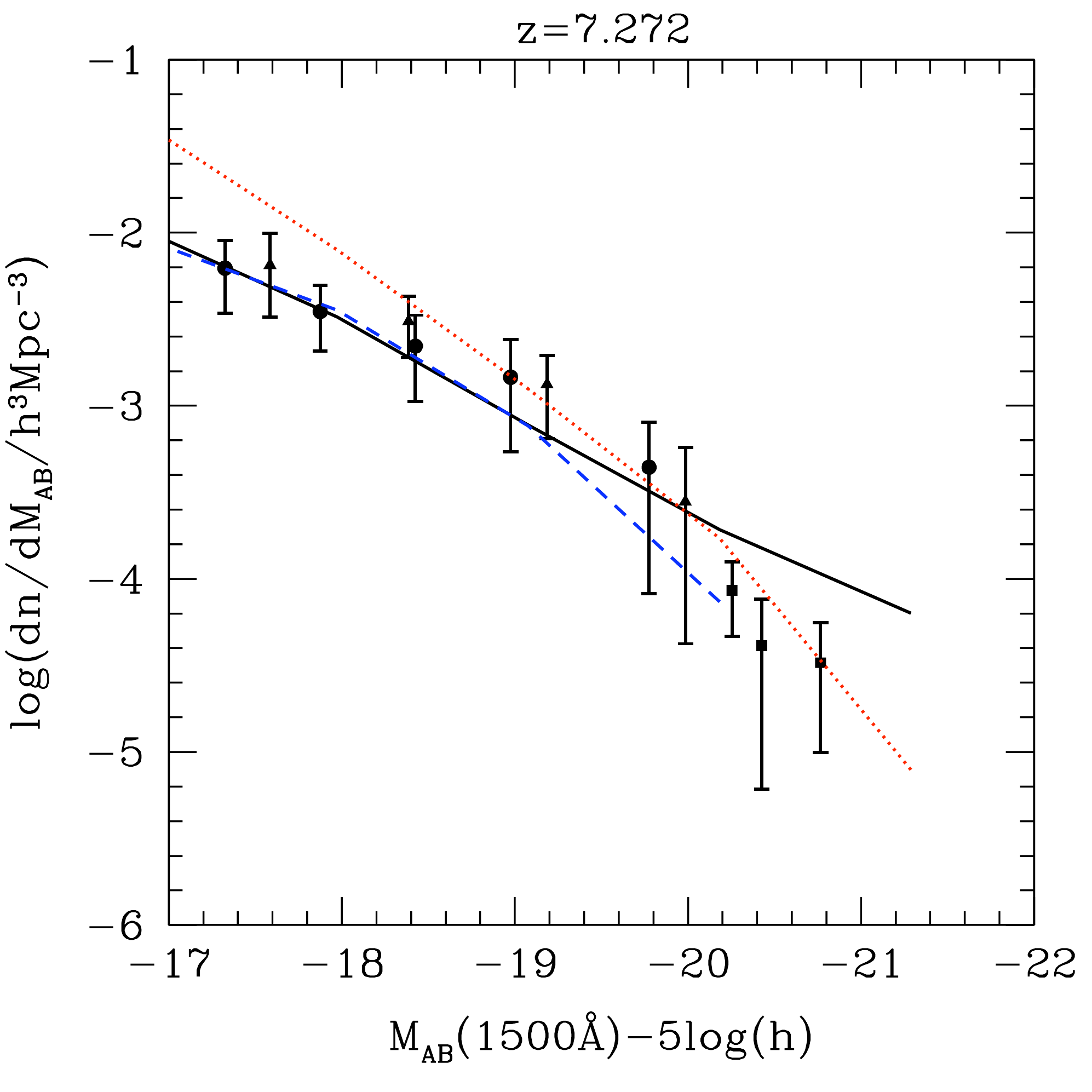}
\includegraphics[width=7cm]{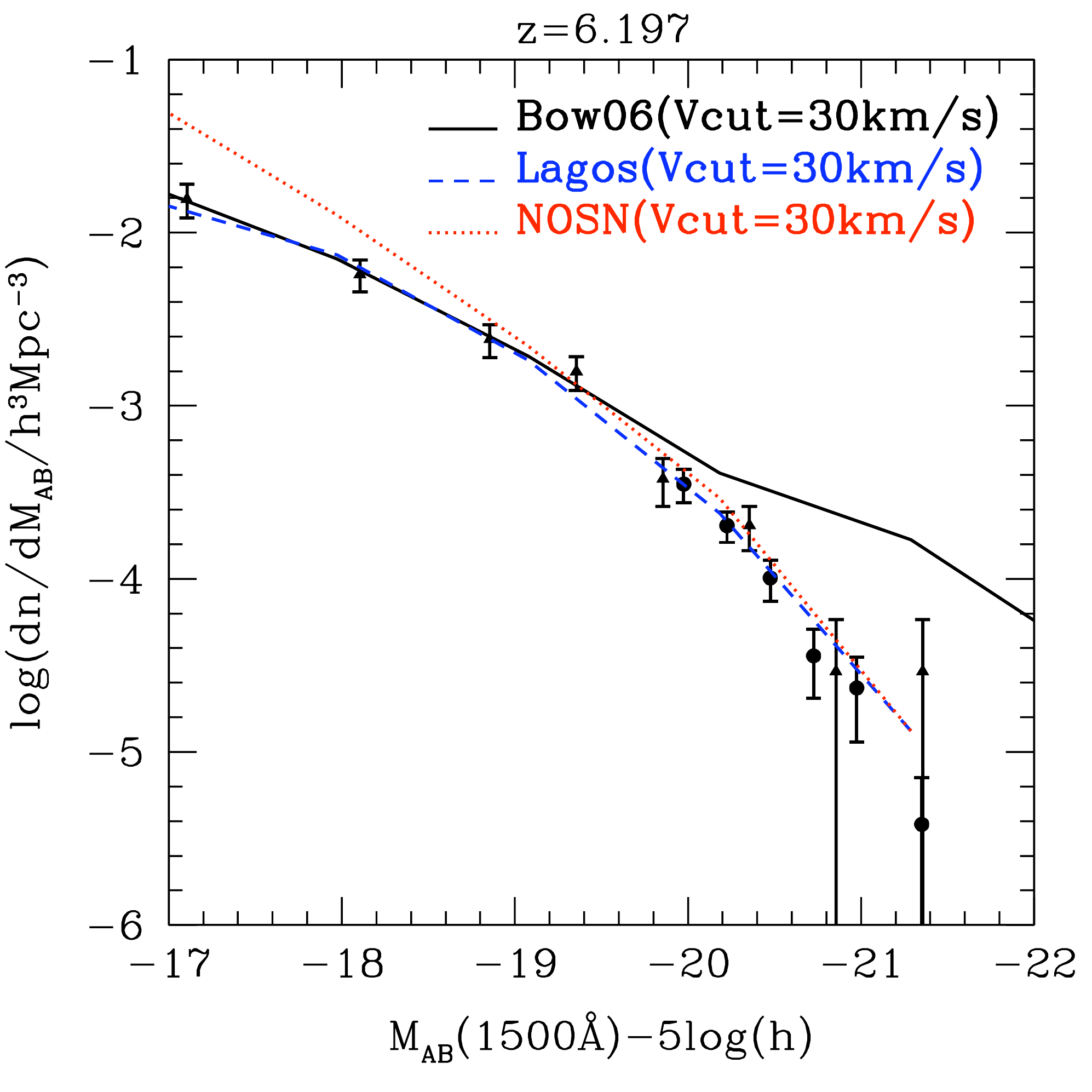}
\end{center}
%\vspace{-8mm}
\caption{The UV luminosity functions from the Bow06, Lagos and NOSN models with observed data points. The top-left panel shows the predicted UV luminosity functions from the models at z=10.074 with the observations estimated by \citet[][]{Bouwens2011} (circle and 1-$\sigma$ upper limit arrows, 1600$\AA$) for $z\sim10$. The top-right panel shows the z=7.883 predictions from the models including the observations for the $z\sim8$ measured by \citet[][]{Bouwens2010} (triangles, 1700$\AA$) and \citet[][]{McLure2010} (circles, 1500$\AA$). The bottom-left is the predictions for z=7.272 from models with the observations for $z\sim7$ came from \citet[][]{McLure2010} (circles, 1500$\AA$), \citet[][]{Oesch2010} (triangles, 1600$\AA$) and \citet[][]{Ouchi2010} (squares, 1500$\AA$). The bottom-right is for z=6.197 predictions with the observatrions for $z\sim6$ measured by \citet[][]{McLure2009} (circles, 1500$\AA$), and \citet[][]{Bouwens2007} (triangles, 1350$\AA$).} \label{UVLF}
\end{figure*}

It is important to consider how well the predicted galaxy population represents the galaxies observed to exist at high redshift. GALFORM models are calibrated to a wealth of data at low redshift.
%Based on this calibration GALFORM can be used to predict the UV luminosity functions (here using the Bow06 model) at high redshift as shown by the solid lines in Figure~\ref{UVLF} for a range of redshifts. For comparison, the Schechter function fits representing the observed luminosity function are also shown as the dotted lines \citep[][]{Bouwens2011}.  Clearly, the Bow06 model significantly under-predicts the number of observed faint galaxies. This difference reduces towards lower redshift. 
Previously, \citet[][]{Lacey2011} used GALFORM with Monte Carlo
merger trees (with no built-in mass resolution limit) to compare model
predictions with observed properties of high-redshift galaxies at $z
\sim 3-10$. Their modelling was successful in reproducing the high redshift luminosity function, although
%Thus, the under prediction of galaxy densities in this work is possibly related to the halo mass resolution of the Millennium Simulation. 
with dependence on the model used. 
%For example, \cite{Lacey2011} show the UVLF of Bow06 and Bau05 models with Monte-Carlo merger trees. 
In our work we have utilised N-body merger trees extracted from the Millennium-II simulation in order to explicitly include the correlations between galaxy position and over-density in the IGM. Since the Millennium-II simulation resolves most of the galaxies responsible for reionization, the model predicts the correct star formation rate density. 
We show the UV luminosity functions at high redshifts in Figure~\ref{UVLF} for the Bow06 model and Lagos model based on the Millennium-II dark matter simulation merger trees. The UV luminosity function using the Millennium-II simulation merger trees is nearly identical to the Monte-Carlo merger trees for the Bow06 model \citep[see ][]{Lacey2011}.
The Bow06 model (solid lines in Fig.~\ref{UVLF}) agrees well with the observational results, with the exception of an over-prediction of luminous galaxies. The Lagos model (dashed lines in Fig.~\ref{UVLF}) shows better agreement with the data across all redshifts and luminosities. The Lagos model has a different star formation law to the Bow06 model, and adopts a burst timescale that also gives better agreement with the UV luminosity function at $z\sim3-7$ following the analysis in the \cite{Lacey2011}. 
%We therefore stress that our modelling is not appropriate for investigating the evolution of the ionization fraction, $\langle x_i\rangle$, with redshift. This shortcoming will be addressed using higher resolution merger trees in the future. However our modelling is sufficient for studying the relative change in the power spectrum due to astrophysical properties in the resolved galaxies. 
 Importantly, both models agree well for the faint galaxies thought to be responsible for reionization, indicating that our results should not be very sensitive to this choice. 
 
 We also show the NOSN model. Simply removing the feedback strength of SNe (by setting $V_{\rm hot}=0$) results in a model that which greatly over predicts the number of galaxies at all luminosities. In order to correct for this we therefore modify the parameter in GALFORM which specifies the ratio between the sum of the mass in visible stars and brown dwarfs, and the mass in visible stars. This parameter ($\Upsilon$) quantifies the assumption for the IMF of brown dwarfs ($m<0.1M_\odot$), which contribute mass but no light to stellar population. We adopt a value of $\Upsilon=4$ for the NOSN and NOSN(no suppression) models. The value of $\Upsilon$ should be greater than unity by definition.

\section{Semi-Numerical sheme to calculate the evolution of ionised structure}
\label{scheme}

%In this section, we show the our method to calculate the 21-cm power spectrum using the hierarchical galaxy formation models. We therefore concentrate on SN feedback effects on the 21-cm power spectrum in the paper.

\citet{MF07} introduced an approximate but efficient method for
simulating the reionization process. This so-called {\em semi-numerical} method extends prior
work by \citet{bond1996a} and \citet{zahn2007}. The method generates an estimate of the ionization field based on a catalogue of sources assigned within the halo field by applying a filtering technique. Good agreement is found with numerical simulations, implying that semi-numerical models can be used to explore a large range of reionization
scenarios. In this paper we apply a semi-numerical technique to find the ionization structure resulting from GALFORM galaxies within the Millennium-II dark matter simulation. 

\subsection{HII regions}
\label{cells}

\begin{figure*}
\includegraphics[width=16cm]{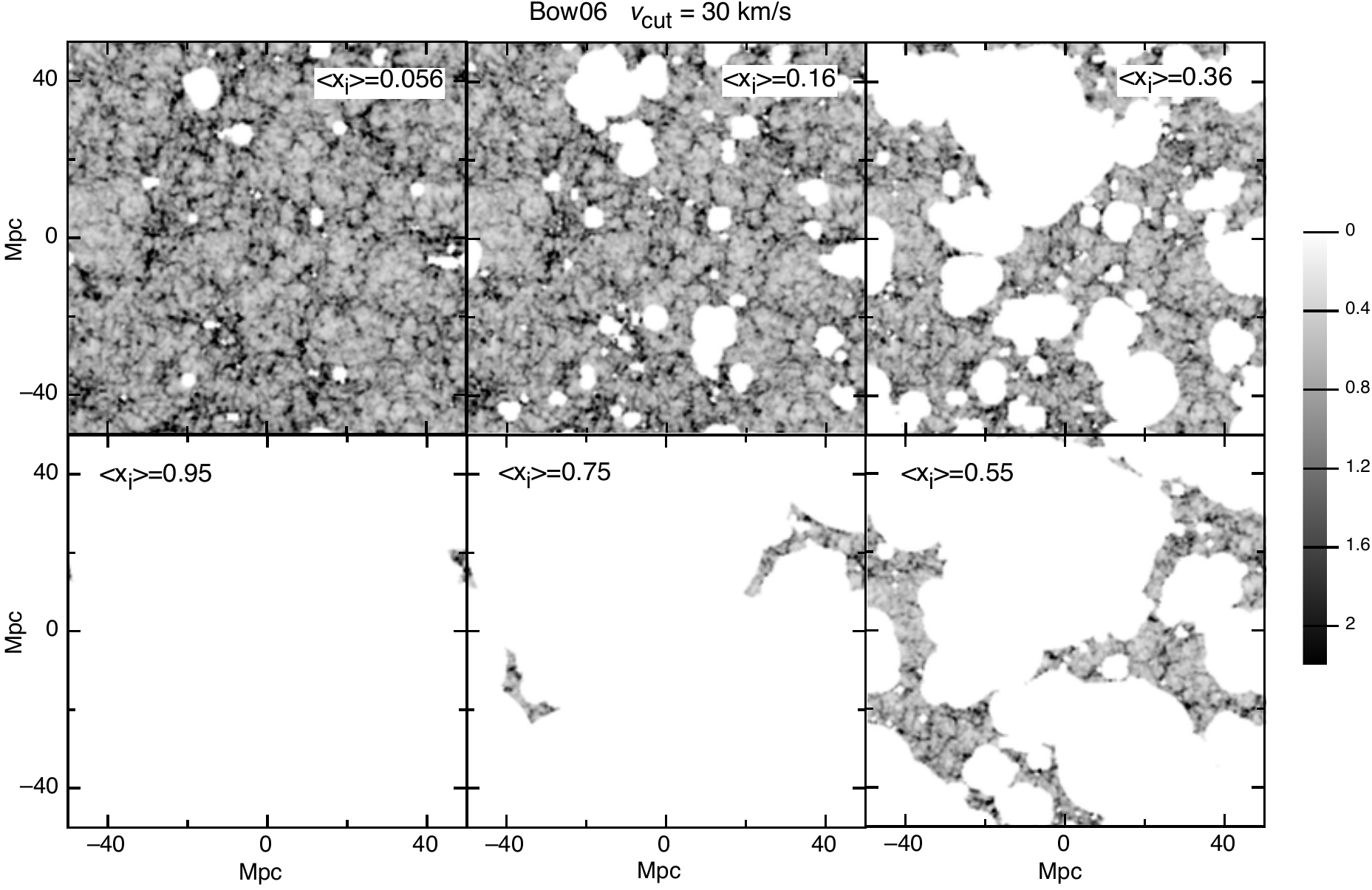}\\
\caption{\label{HIIevolve1}Maps of the 21~cm intensity in slices for a range of values of $\left<x_{i}\right>$ corresponding to different stages of reionization. We assume the Bow06 model. The units of the grey-scale are $(28[(1+z)/10]$mK$)$. The slices are 0.3906$h^{-1}$Mpc deep.}
\end{figure*}

\begin{figure*}
\begin{center}
\includegraphics[width=16cm]{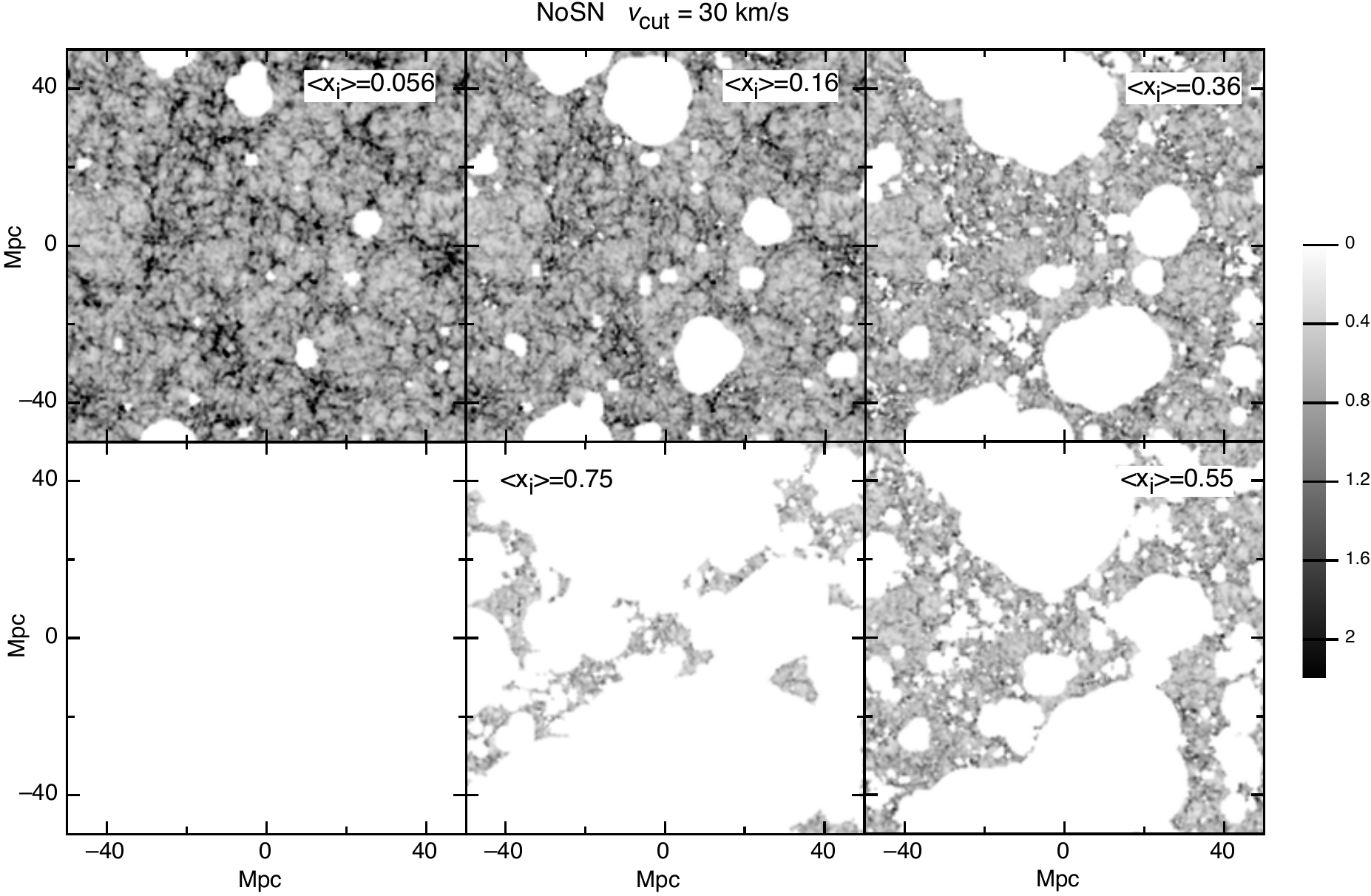}
\end{center}
\caption{\label{HIIevolve2}Maps of the 21~cm intensity in slices for a range of values of $\left<x_{i}\right>$ corresponding to different stages of reionization. We assume the NOSN model. The units of the grey-scale are $(28[(1+z)/10]$mK$)$. The slices are 0.3906$h^{-1}$Mpc deep.}
\vspace{5mm}
\begin{center}
\includegraphics[width=16cm]{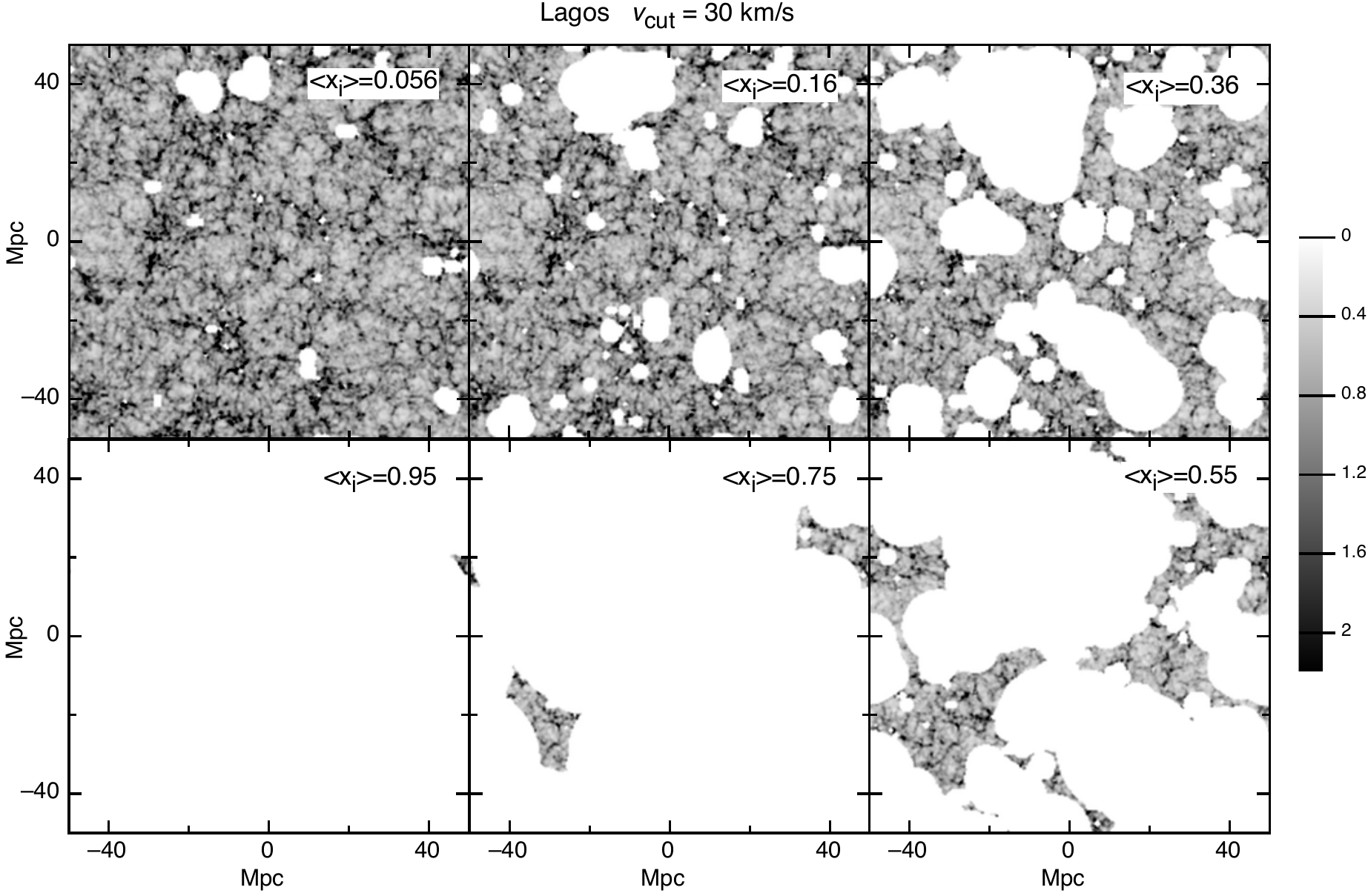}
\end{center}
\caption{\label{HIIevolve3}Maps of the 21~cm intensity in slices for a range of values of $\left<x_{i}\right>$ corresponding to different stages of reionization. We assume the Lagos model. The units of the grey-scale are $(28[(1+z)/10]$mK$)$. The slices are 0.3906$h^{-1}$Mpc deep.}
\end{figure*}

\begin{figure*}
\begin{center}
\includegraphics[width=16cm]{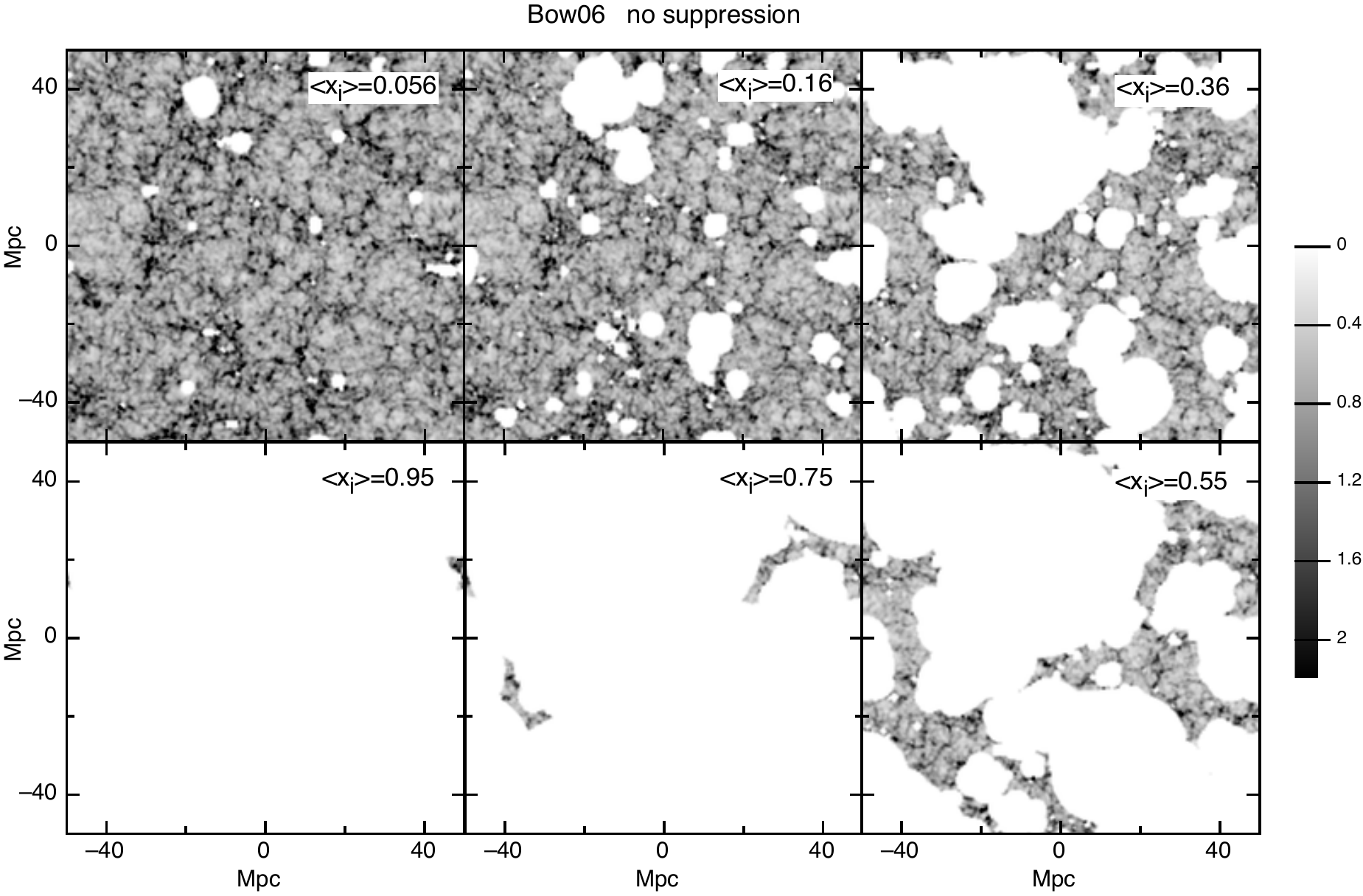}
\end{center}
\caption{\label{HIIevolveII4}Ionization maps for a range of values of $\left<x_{i}\right>$ corresponding to different stages of reionization. We assume the Bow06(no suppression) model. The units of the grey-scale are $(28[(1+z)/10]$mK$)$. The slices are 0.3906$h^{-1}$Mpc deep.}
\begin{center}
\vspace{5mm}
\includegraphics[width=16cm]{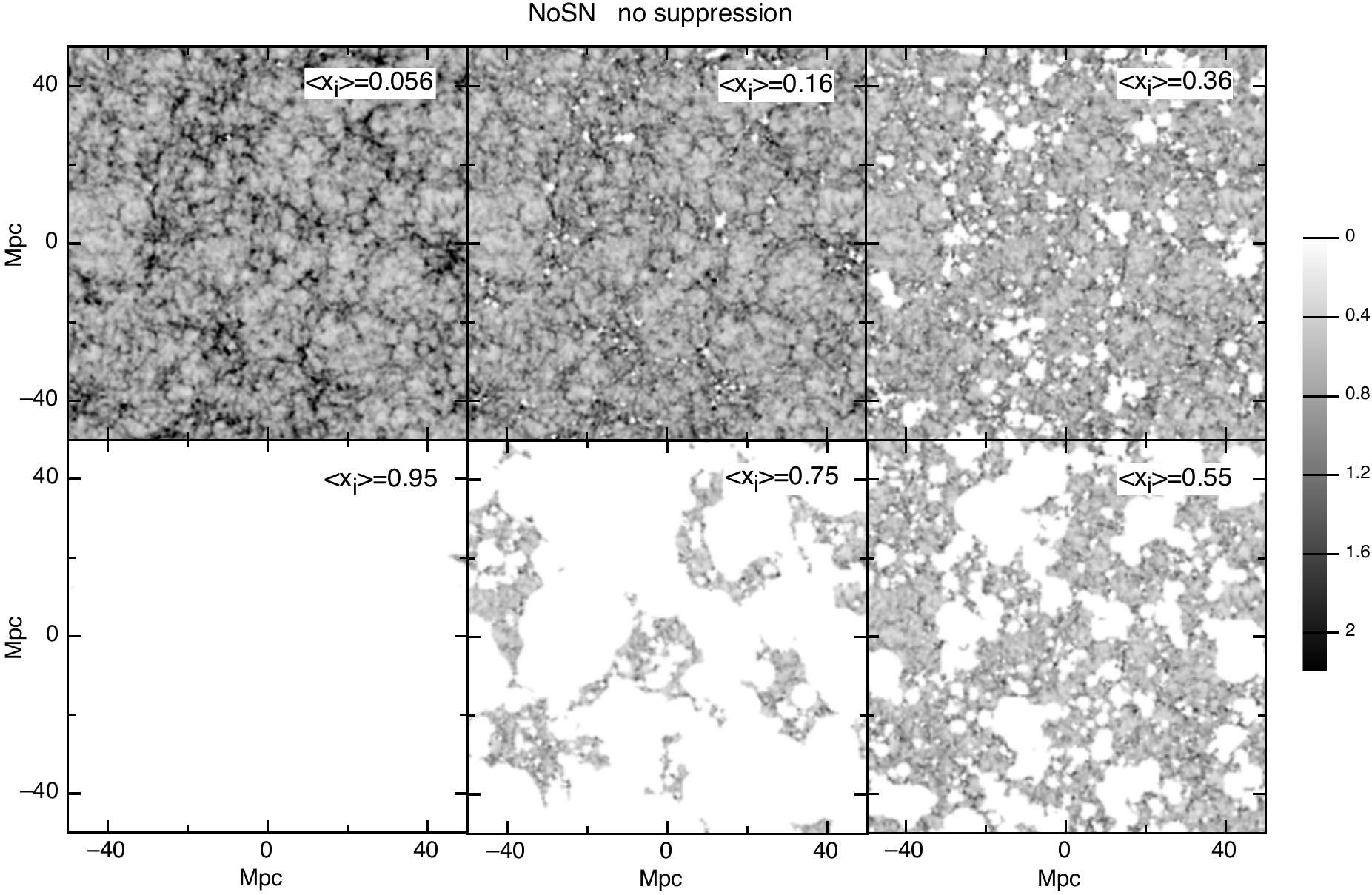}
\end{center}
\caption{\label{HIIevolveII5}
Ionization maps for a range of values of $\left<x_{i}\right>$ corresponding to different stages of reionization. We assume the NOSN(no suppression) model. The units of the grey-scale are $(28[(1+z)/10]$mK$)$. The slices are 0.3906$h^{-1}$Mpc deep.}
\end{figure*}

\begin{figure*}
\begin{center}
\includegraphics[width=17cm]{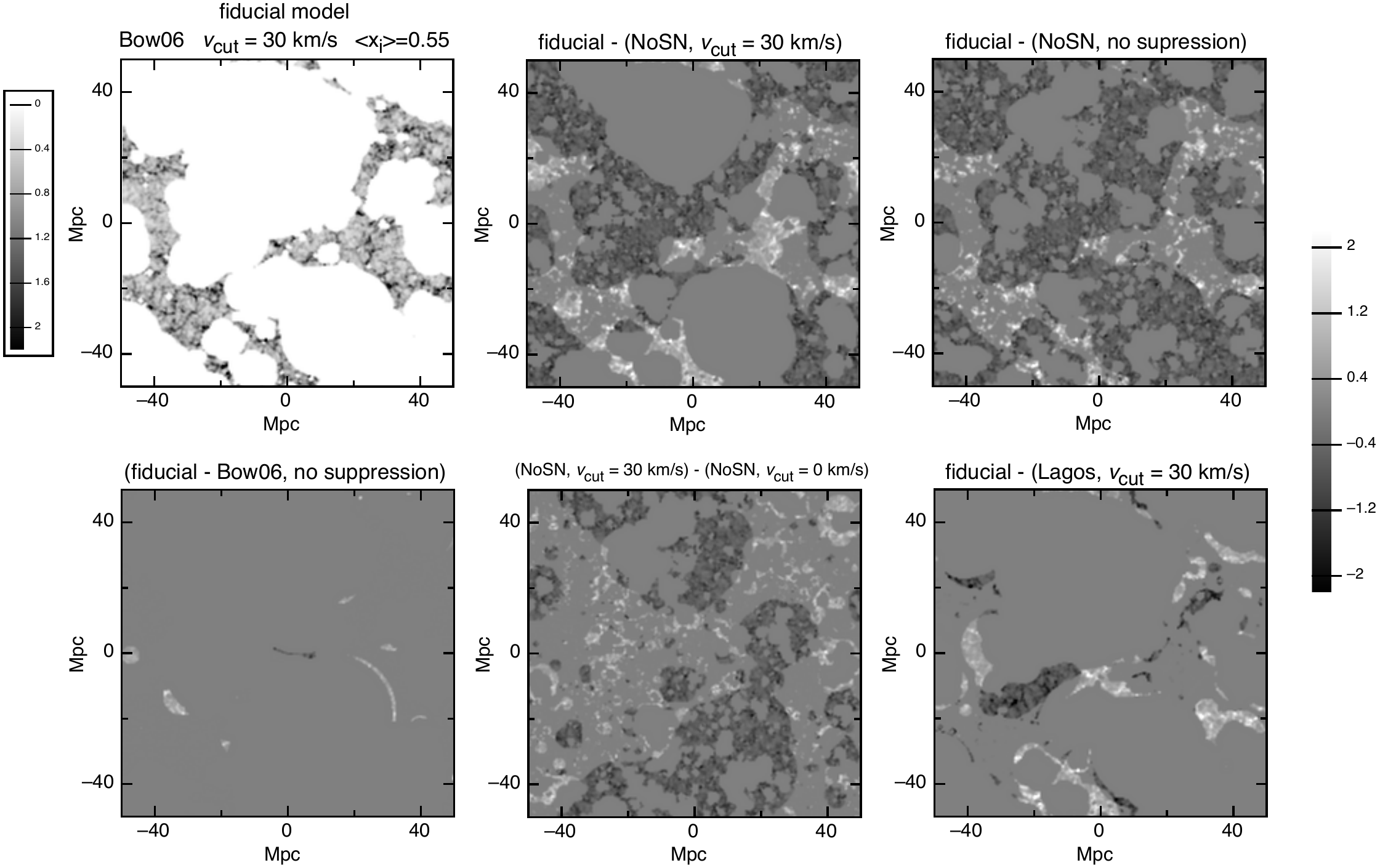}
\end{center}
\vspace{-3mm}
\caption{\label{HIIB}Example maps of the ionization structure produced by our modelling. In each case the slices shown are 100h$^{-1}$Mpc on a side, and 0.3906h$^{-1}$Mpc deep. The models were computed with $\left<x_{i}\right>=0.55$ ($z=7.272$). The top-left panel shows map for the Bow06 model. To make the effect of SNe and photo-ionization feedback on the ionization map clear, the others show the subtraction of the Bow06 map from the variant model's maps except the bottom-middle panel. Here a positive value shows regions where the Bow06 model predicts HII region but variant models does not. Conversely the negative values show regions where the variant models predict ionization but the Bow06 model does not. The units of the grey-scale are $(28[(1+z)/10]$mK$)$.}\
%The blue rings illustrate the increased size of HII regions in the Bow06 model compared with the NOSN model.
\end{figure*}

We begin by binning galaxies from the GALFORM model into small regions of volume (or cells). We assume the number of photons produced by galaxies in the cell that enter the IGM and participate in reionization to be
\begin{equation}\label{nphotons}
N_{\rm \gamma, cell}={\it f}_{\rm esc}\int^{t_{z}}_{0}\dot{N}_{\rm Lyc,cell}(t)~dt,
%N_{\rm \gamma, cell}=N_{\rm photon}(\rm{IMF},{\it Z}) {\it f}_{\rm esc} {{{\it M}_{\rm \star,cell}} \over {\it m}_{\rm p}},
\end{equation}
where $f_{\rm esc}$ is the escape fraction of photons produced by stars in a galaxy. The total Lyman continuum luminosity of the $N_{\rm cell}$ galaxies within the cell expressed as the emission rate of ionizing photons { (i.e.  units of photons/s)} computed from GALFORM is
\begin{equation}
\dot{N}_{\rm Lyc,cell}(t) = \sum_{i=1}^{N_{\rm cell}} \dot{N}_{\rm Lyc, \it i}(t),
\end{equation}
where
\begin{equation}
\dot{N}_{\rm Lyc, \it i}(t)=\int^{\infty}_{\nu_{\rm thresh}}{L_{\nu,i}(t) \over h\nu} {\rm d\nu},
\end{equation}
L$_{\nu,i}$ is the spectral energy distribution of galaxy $i$ and $\nu_{\rm thresh}$ is the Lyman-limit frequency, $h\nu_{\rm thresh}$ = 13.6 eV.
Note that the number of photons produced per baryon in long-lived stars and stellar remnants depends on the IMF and metalicity ({\it Z}).
The stellar population models used in GALFORM output give
4.54$\times$10$^{3}$ for {\it Z}=0.02 and 6.77$\times$10$^{3}$ for {\it Z}=0.004
using the Kennicutt IMF. { Note that we assume the total Lyman continuum luminosity in a cell at redshift $z_{i}$ to be constant until the next snapshot at redshift $z_{i+1}$, and calculate the number of photons produced in the cell between $z_{i}$ and $z_{i+1}$ 
to be $\dot{N}_{\rm Lyc,cell}(t_{z_{i}})  \times (t_{z_{i+1}}-t_{z_{i}})$.}

We then calculate the ionization fraction within each cell according to  
\begin{equation}
\label{Qvalue}
Q_{\rm cell}=\left[{N_{\rm \gamma, cell} \over (1+F_{\rm c})N_{\rm HI, cell}}\right],
\end{equation}
where $F_{\rm c}$ denotes the mean number of recombinations per hydrogen atom up to reionization and $N_{\rm HI, cell}$ is the number of neutral hydrogen atoms within a cell. The latter quantity is calculated as 
\begin{equation}
\label{nHI}
N_{\rm HI, cell}={n_{\rm HI}(\delta_{\rm DM,cell}+1)V_{\rm cell}},
\end{equation}
where we assume that the overdensity of neutral hydrogen follows the dark matter (computed based on the Millennium-II simulation density field), $n_{\rm HI}$ is the mean comoving number density of hydrogen atoms, and $V_{\rm cell}$ is the comoving volume of the cell. Self-reionization of a cell occurs when $Q_{\rm cell}=1$. We divide the Millennium-II simulation box into 256$^{3}$ cells, yielding cell side lengths of 0.3906$h^{-1}$Mpc and comoving volumes of 0.0596$h^{-3}$Mpc$^{3}$.  

Theoretical prediction of the parameters $F_{\rm c}$ and $f_{\rm esc}$ in Equations~(\ref{nphotons}) and~(\ref{Qvalue}) is complicated, and their values are not known. The recombination parameter $F_{\rm c}$ is related to the density of the IGM on small scales, while $f_{\rm esc}$ depends on the details of the high redshift ISM. Previous work using GALFORM suggested the value $(1+F_{\rm c})/f_{\rm esc}\sim10$ \cite[][]{Benson2001,Theuns2011} to fit observational constraints on reionization. Here, we determine the value of $(1+F_{\rm c})/f_{\rm esc}$ that is required in order to give a particular value of ionization fraction at each redshift. We explicitly note here our assumption that values of $(1+F_{\rm c})/f_{\rm esc}$ do not depend on galaxy mass. In reality the escape  fraction may be mass dependent, particularly in models with SNe feedback. 

Based on equation~(\ref{Qvalue}), individual cells can have $Q_{\rm cell}>1$. On the other hand, cells with $Q_{\rm cell}<1$ may be ionized by photons produced in a neighbouring cell. In order to find the extent of ionized regions we therefore filter the $Q_{\rm cell}$ field using a sequence of real space top hat filters of radius $R$ (with $0.3906<R<100h^{-1}$Mpc), producing one smoothed ionization field $Q_R$ per radius. At each point in the simulation box we find the largest $R$ for which the filtered ionization field is greater that unity (i.e. ionized with $Q_R>1$). All points within the radius $R$ around this point are considered ionized. This procedure forms the position dependent ionization fraction $0\leq Q\leq1$, which describes the ionization structure of the IGM during reionization. The filtering follows the method outlined in more detail in \citet[][]{GW08}.

%Thus in our models the feedback processes associated with AGN and photoionization affect the predicted 21cm power spectrum less than SNee feedback. In the case of AGN, this is because there are few massive haloes at {\it z}$\sim$6 which are affected by AGN feedback. On the other hand in the case of photoionization feedback, we are unable to investigate the effect since the resolution of the Millennium-II simulation does not include haloes which are affected by the photoionization feedback. 

%The most striking trend is the predicted difference of the 21-cm power spectrum for models with different SN feedback strength (i.e. the Bow06, 300SN, NOSN models). The lower panel in Figure~\ref{PSModels} shows the fractional difference between these models and the NOSN model. The NOSN model has a smaller amplitude over a broad range of spatial scales than do those models that include SNe feedback with the differences being scale dependent. Moreover, the amplitude of the power spectrum increases with increasing SN feedback strength. 

%The Bow06 model has more blue points, Q$_{cell}$ value greater than 0 and less than 1, outside the HII region than NOSN model. This means that the total number of cell and shape of HII region is different even though the mass averaged ionization fraction, $\left<x_{i}\right>$, is same. Therefore the predicted 21-cm power spectrums from the models show the difference.
 
\subsection{redshifted 21cm intensity} 
 
Fluctuations in 21~cm intensity (or brightness temperature) from different regions of the IGM include contributions from a range of different physical properties, including density, velocity gradients, gas temperature, hydrogen spin temperature and ionization state \citep{FOB06}. In this paper we restrict our attention to analyses that assume the spin temperature of hydrogen is coupled to the kinetic temperature of an IGM that has been heated well above the CMB temperature (i.e. $T_{\rm s}\gg T_{\rm CMB}$). This condition should hold during the later stages of the reionization era \citep[$z\la9$, ][]{S+07}. In this regime there is a proportionality between the ionization fraction and 21~cm intensity, and the 21~cm brightness temperature contrast may be written
\begin{equation}
\label{Tb}
\Delta T=23.8\left(\frac{1+z}{10}\right)^\frac{1}{2}\left[1-Q\right]\left(1+\delta_{\rm DM,cell}\right)\,\mbox{mK}.
\end{equation}
Here we have ignored the contribution to the amplitude from velocity gradients, and assumed as before that the hydrogen overdensity follows the dark matter ($\delta_{\rm DM,cell}$).

 \begin{table*}
\caption{
The values of  $(1+F_{\rm c})/f_{\rm esc}$ corresponding to the different models and redshifts shown in this paper.}
\label{Fraction}
\begin{tabular}{ccccccc}
\hline
\hline
Redshift (z) & 9.278&8.550&7.883&7.272&6.712&6.197\\
$\left<x_{i}\right>$ & 0.056&0.16&0.36&0.55&0.75&0.95\\
\hline
Bow06  & 4.85& 3.24 &2.72& 3.10& 3.83&4.82\\
Bow06(no suppression) &4.856  & 3.24 &2.71 &3.12 &3.85 &4.81 \\
Lagos & 3.86& 2.61&2.17 & 2.53&3.11 &3.95 \\
NOSN & 417.98&189.28&106.70&85.59 &74.83&68.94 \\
NOSN(no suppression) & 267.78& 136.97& 85.03&73.86 &69.62 &68.28\\
\hline
\end{tabular}
\end{table*}

\section{Structure of Reionization}
\label{Maps}

In this section we present results for the possible structure of ionisation in the IGM. Figures~\ref{HIIevolve1} - \ref{HIIevolveII5}, we show example ionization maps for our five different models. In each case we show examples for $\left<x_{i}\right> = $ 0.95, 0.75, 0.55, 0.36, 0.16 and 0.056, illustrating the growth of HII regions during reionization. For our model these values correspond to redshifts of $z\sim$ 6.197, 6.712, 7.272, 7.883, 8.550 and 9.278 [selected for comparison with the work by \cite{lidz2008}]. Maps are shown for Bow06, NOSN and Lagos models in Figure~\ref{HIIevolve1} - \ref{HIIevolve3}, and for Bow06(no suppression) and NOSN(no suppression) models in Figure~\ref{HIIevolveII4} - \ref{HIIevolveII5}. In order to make this comparison we adjust the quantity $(1+F_{c})/f_{\rm esc}$ in equations~(\ref{nphotons}) and~(\ref{Qvalue}) so as to get same mass averaged ionization fraction $\left<x_{i}\right>$ for all models at each redshift. The values of $(1+F_{\rm c})/f_{\rm esc}$ required in order to give a particular values of ionization fraction at each redshift are shown in Table~\ref{Fraction}. Models presented in this work take values of $(1+F_{\rm c})/f_{\rm esc}$ that are less than 10 for models including SNe feedback in agreement with the work of \cite{Benson2001} and \cite{Theuns2011}, but greater than 50 for models without SNe feedback. 

The effects of SNe feedback strength (between the Bow06 model and the NOSN model) and of star formation law (between the Bow06 model and the Lagos model) on the ionization history can be seen explicitly by comparing Figures~\ref{HIIevolve2} and \ref{HIIevolve3} with Figure~\ref{HIIevolve1}. The regulation of star formation and cooling of hot gas in small galaxies by the SNe feedback process leads to massive galaxies which are more biased towards dense regions, dominating the production of ionizing photons. As a result, the evolution of large HII regions in the Bow06 model (Figures~\ref{HIIevolve1}) starts from the overdense environment and propagates to neighbouring overdense regions. Conversely, the production of ionization photons from the massive galaxies is less prominent in the NOSN model (Figures~\ref{HIIevolve2}) than in the Bow06 model. As a result the HII region evolution maps for the NOSN model show many more smaller HII regions than in the Bow06 case. The different star formation law prescriptions in the Lagos (Figures~\ref{HIIevolve3}) and Bow06 models also lead to differences in evolution of HII regions. However the variation is much smaller than is found from differences in the SNe feedback strength.
 
The effect of photo-ionization feedback on the HII regions evolution can be seen by comparing Figures~\ref{HIIevolveII4} and \ref{HIIevolveII5} with Figure~\ref{HIIevolve1}. We find very little difference between the HII region evolution in the Bow06 model and the Bow06(no suppression) model (Figure~\ref{HIIevolveII4}). Thus the effect of SNe feedback on the evolution of ionisation structure is much larger than that from photo-ionization feedback. However in the absence of SNe feedback, the effect of photo-ionization feedback effect is significant (Figure~\ref{HIIevolveII5}), and the NOSN(no suppression) model produces numerous, smaller HII regions that are relatively homogeneously distributed through the IGM.

To highlight the differences between the maps produced by models with and without SNe and/or photo-ionization feedback, in Figure~\ref{HIIB}  we show differences between ionisation maps for the Bow06 and other models. The exception is the lower-middle panel, which shows the difference between maps for the NOSN and NOSN(no suppression) models. In all cases the mass averaged ionization fraction is $\left<x_{i}\right>=0.55$ $(z=7.272)$. Positive  values represent the area where the Bow06 model predicts HII regions but alternative models  do not (in the lower-middle panel the positive values represent the area where the NOSN model predicts HII regions but the NOSN(no suppression) does not). This figure clearly shows the large effect of SNe feedback relative to photo-ionization feedback and star formation prescription.

\section{Contributions to the ionising photon budget}
\label{budget}

\begin{table}
\caption{
The values of  $F_{N_{\rm photons}}$, the fraction of ionizing photons where the galaxy circular velocity less than 30km/s.}
\label{Npfrac}
\begin{center}
\begin{tabular}{ccc}
\hline
\hline
Model & $V_{\rm halo}< $30Km/s& All\\
\hline
Bow06(no suppression) & 0.0028 & 1\\
NOSN(no suppression)  & 0.39&1 \\
\hline
\end{tabular}
\end{center}
\end{table}

Radiative feedback has been thought to play a significant role in self-regulating the reionization process by suppressing galaxy formation in reionized regions \citep[e.g.][]{iliev2007}. These studies were based on the assumption that the ionising luminosity to halo mass ratio does not depend on halo mass. However the presence of SNe feedback in galaxy formation models is known to modify the mass-to-light ratio of galaxies through regulation of star formation. Motivated by the unexpectedly small difference in ionising structure between models with SNe feedback that do and do not include radiative feedback, in this section we calculate the effect of radiative feedback on stellar mass and ionising photon contribution.

\begin{figure}
\begin{center}
\includegraphics[width=8.5cm]{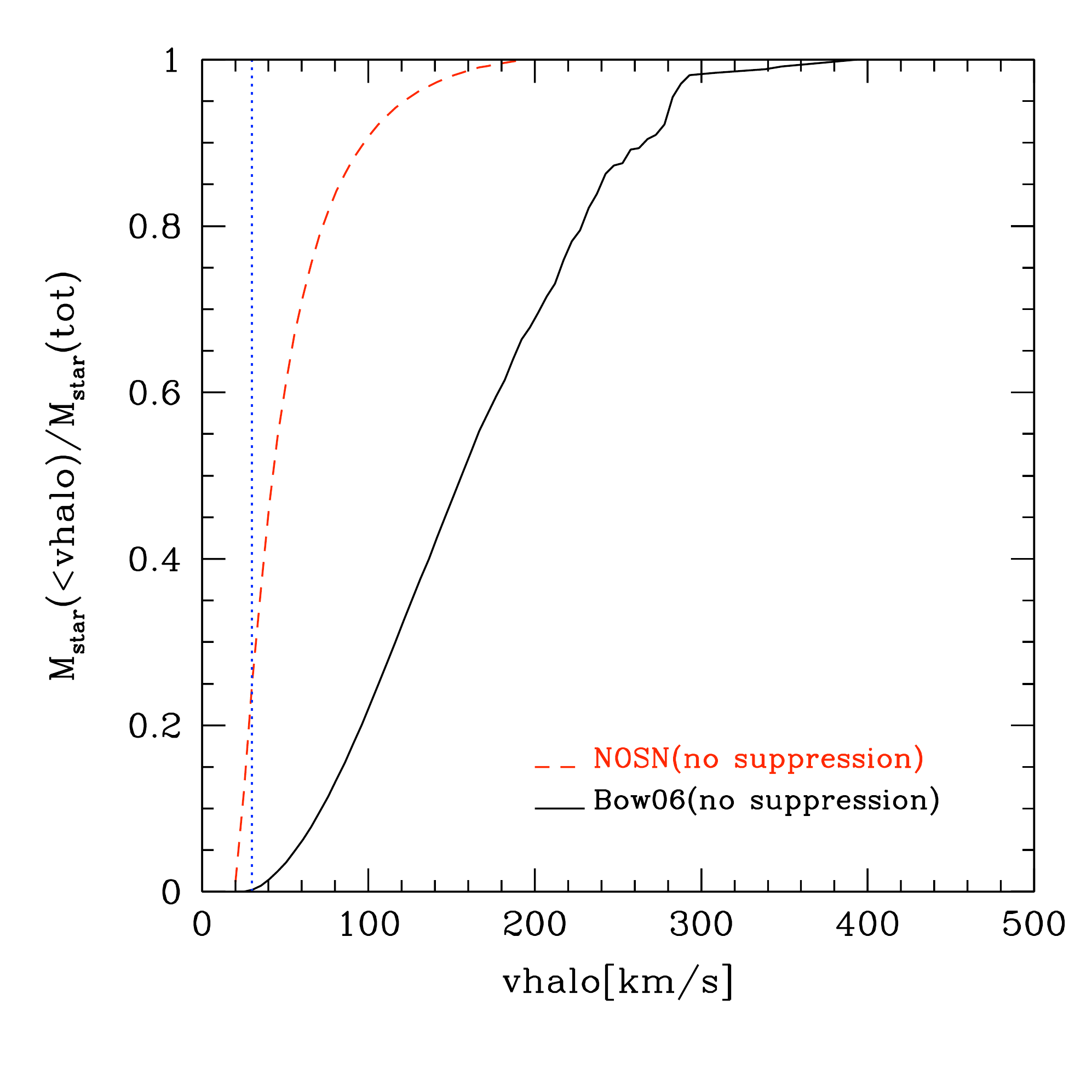}
\end{center}
\vspace{-8mm}
\caption{The cumulative fraction of stellar mass as a function of $V_{\rm halo}$[km/s] at $\left<x_{i}\right>=0.55$ (corresponding to $z=7.272$) for the NOSN(no suppression) and Bow06(no suppression) models. The vertical line indicates the value of $V_{\rm cut}=30$km/s.} \label{DisDQ}
\end{figure}

First, in Figure~\ref{DisDQ}, we show the cumulative fraction of stellar mass as a function of $V_{\rm halo}$[km/s] at $\left<x_{i}\right>=0.55$ for the Bow06(no suppression) and the NOSN(no suppression) models. The contribution to total stellar mass from galaxies which have circular velocities $V_{\rm halo}$ smaller than 30km/s (left dotted line) is almost zero for the Bow06(no suppression) model and 25 percent for the NOSN(no suppression) model. This shows that SNe feedback greatly lowers the potential contribution of low circular velocity galaxies, which are the ones affected by the photo-ionization feedback process, and explains why SNe feedback is the more dominant effect. 

Similarly, we also calculate the fraction $F_{N_{\rm photons}}$ of ionising photons produced by galaxies with $V_{\rm halo}$ less than 30km/s for the Bow06(no suppression) and the NOSN(no suppression) models. The values are listed in Table~\ref{Npfrac}, and show the small fraction of photons  produced by the low mass haloes which are subject to the radiative feedback process; see also \citet[][]{Theuns2011} for similar discussion. Thus we find that SNe feedback renders the effect of radiative feedback on the reionization history negligible, indicating that reionization is not self regulating, as indicated in previous work \citep[e.g.][]{iliev2007}.

{ While the calculations presented above provide the quantitative estimate of the effect of SNe feedback on the photon budget, we can also provide a simple argument to show qualitatively why SNe feedback should be  the dominant process governing the contribution of low mass galaxies to reionization \citep[e.g][]{Benson2006, Theuns2011}. Ignoring photo-ionization feedback ($V_{\rm cut}$=0), gas cooling is very efficient in halos with a virial temperature of $T_{\rm vir}\sim10^{4}$-$10^{5}$K. We assume that the resulting star formation timescale is shorter than the Hubble time, in which case all of the cooled gas will either form stars or be ejected by SNe feedback (equation~\ref{snbeta}), yielding 
\begin{equation}
(1-R+\beta)M_{\rm \star}\sim M_{\rm b},
\end{equation}
where $M_{\rm \star}$ is the mass of stars formed (before recycling), $0<R<1$ is the recycled fraction, and $M_{\rm b}$ is the total mass of 
baryons in a halo of mass $M_{\rm halo}$. For low mass galaxies (with $V_{\rm cut}\sim30$km/s) $\beta\gg1$, and the fraction of baryons converted into stars is $M_{\rm \star}/M_{\rm b}\sim1/\beta\sim 10^{-3}(V_{\rm halo}/30\mbox{km\,s}^{-1})^3 \propto M_{\rm halo}$. The total ionising contribution to reionization is proportional to the product of this fraction and the mass in dark matter halos (i.e. $(M_{\rm \star}/M_{\rm b})M_{\rm halo}\propto M_{\rm halo}^2$). At low masses the halo mass function (number density per unit mass) is $dn/dM_{\rm halo}\propto M_{\rm halo}^{-\gamma}$ with $\gamma \approx 2$. The mass in stars per logarithm of halo mass per unit volume in the Universe is therefore proportional to $ M_{\rm halo}^2\times M_{\rm halo}dn/dM_{\rm halo}\propto M_{\rm halo}$. Thus, we find that very low mass galaxies should contribute little to reionization \citep[see also][]{Stuart2012}.}

\section{The 21-cm power spectrum}
\label{PS}

The filtering procedure described in \S~\ref{scheme} provides a 3-dimensional map of the ionization structure within the Millennium-II Simulation box, which provides a 3 dimensional 21cm intensity cube via equation~(\ref{Tb}). From this cube we calculate the dimensionless 21-cm power spectrum 
\begin{equation}
\Delta^{2}(k)=k^3/(2\pi^2)P_{21}(k)
\end{equation}
as a function of spatial frequency $k$, where $P_{21}(k)$ is the 21cm power spectrum. When calculating the power spectrum, velocity gradients increase the amplitude of the spherically averaged redshift space power spectrum by a factor of $4/3$ relative to the real space power spectrum based on linear theory \citep[][]{bl2005}. { \cite{Mao2012} shows that this factor can be much higher than $4/3$ over the intermediate range $k\sim0.1-1 {h/ \rm Mpc}$ at the epoch where the IGM is 50$\%$ ionized.}

%\begin{figure}
%\begin{center}
%\includegraphics[width=8.5cm]{PKLFMIIN}
%\end{center}
%\vspace{-8mm}
%\caption{The predicted dimensionless 21-cm power spectra for the models discussed in this paper. The models are each shown at the same value of mass averaged ionization fraction $\left<x_{i}\right>=0.55$ ($z=7.272$). The units of the power spectrum are (28[(1+z)/10]mK)$^{2}$. The lower panel shows the ratios of the four models to the Bow06 model.  } \label{PSModels}
%\end{figure}

\begin{figure*}
\begin{center}
\includegraphics[width=5.3cm]{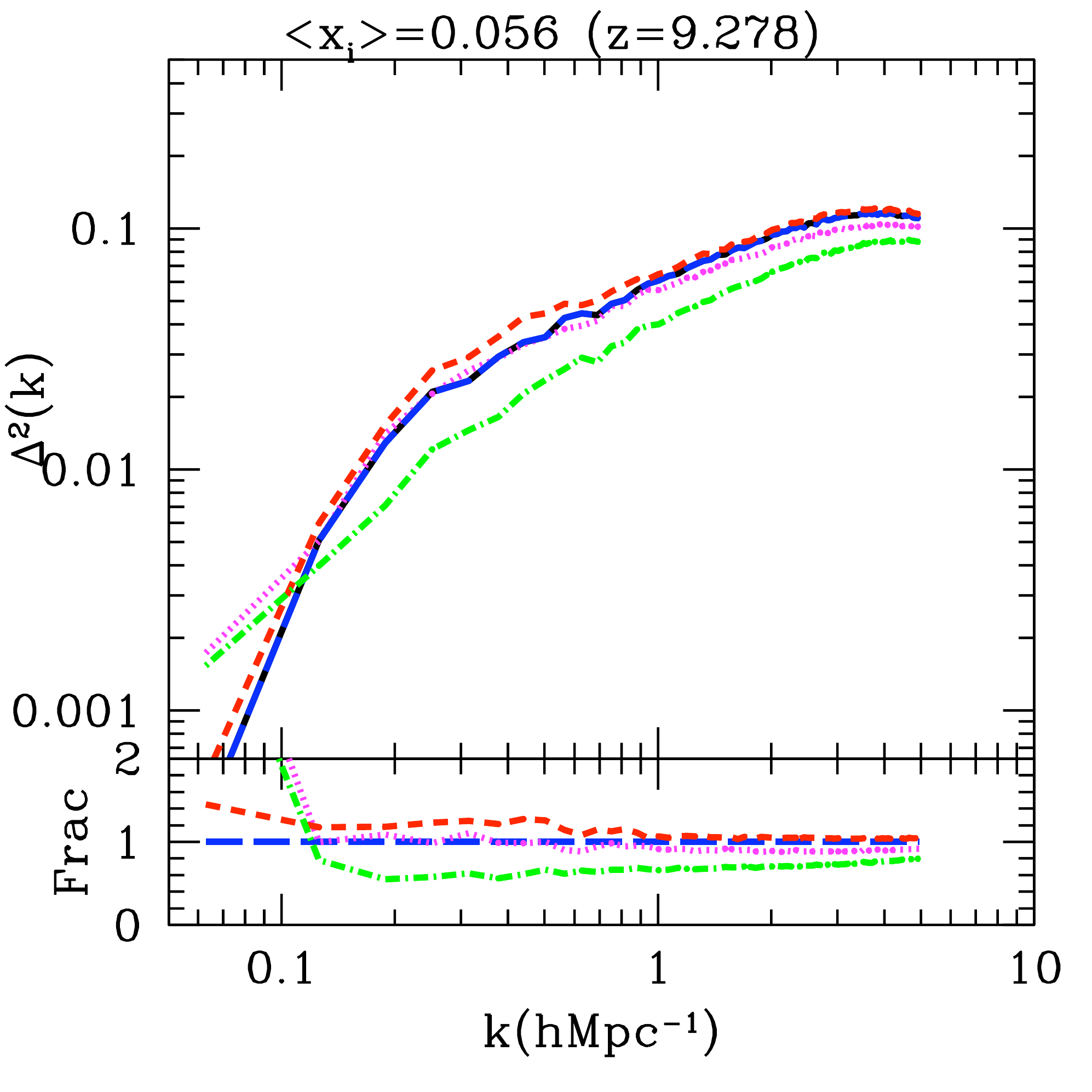}
\includegraphics[width=5.3cm]{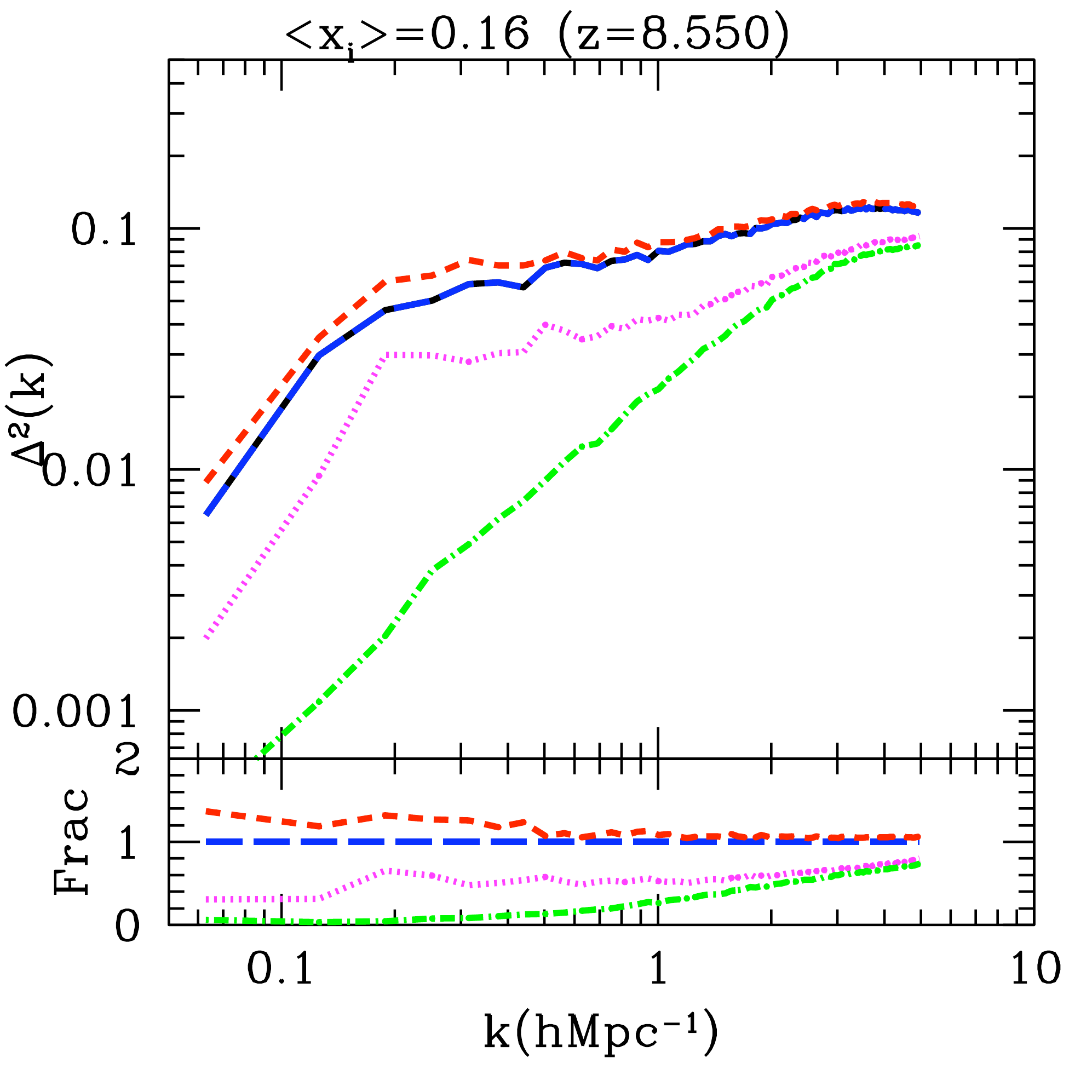}
\includegraphics[width=5.3cm]{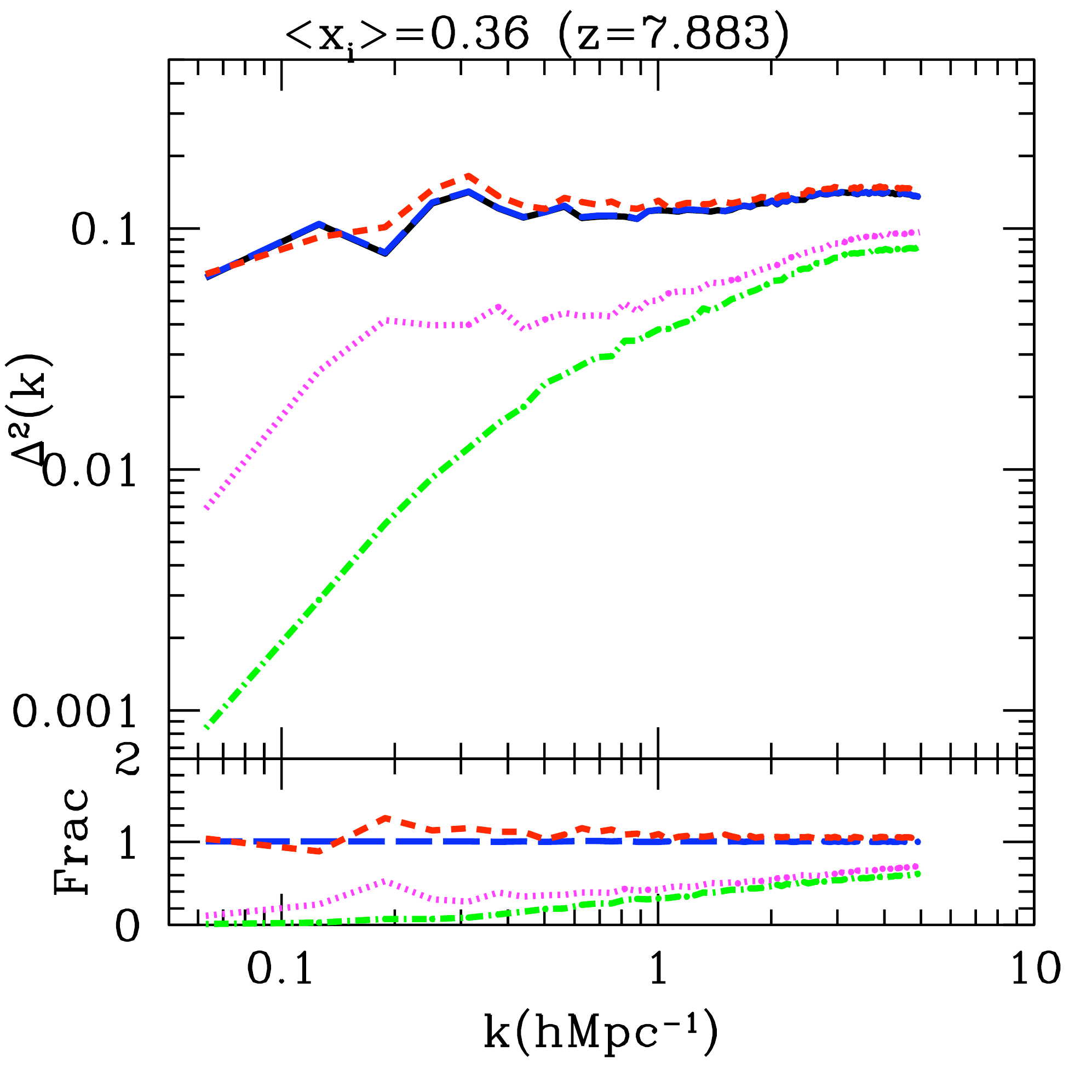}
\includegraphics[width=5.3cm]{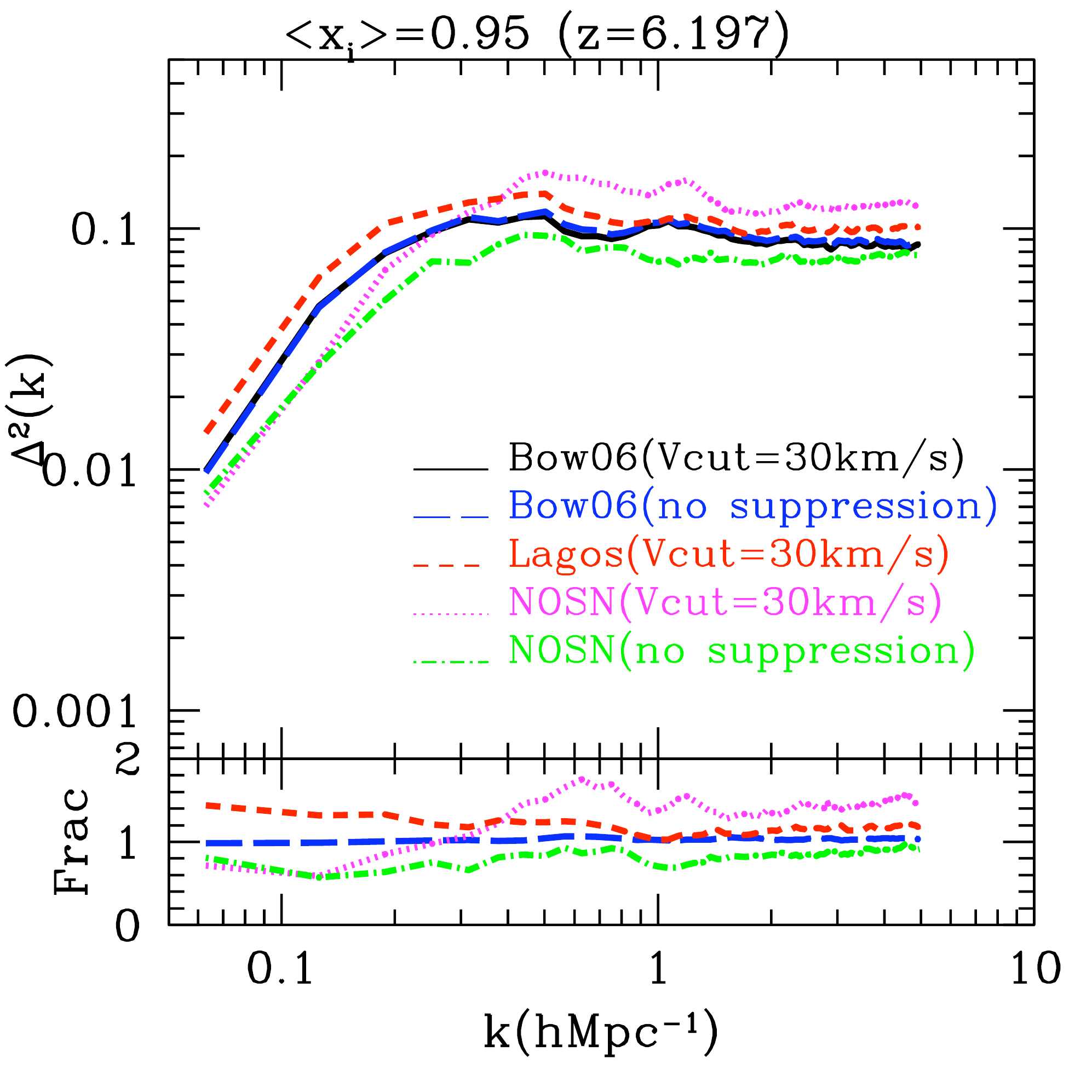}
\includegraphics[width=5.3cm]{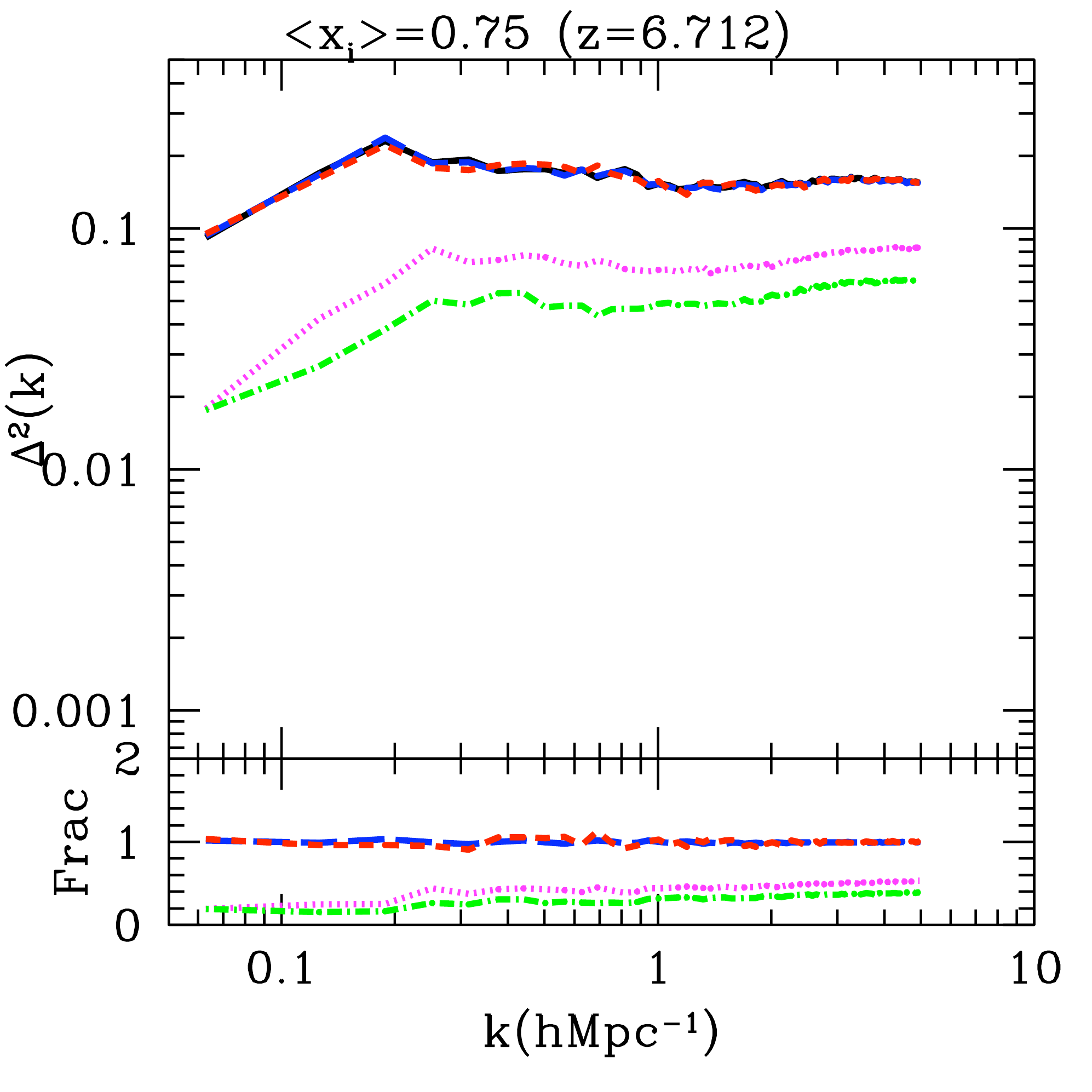}
\includegraphics[width=5.3cm]{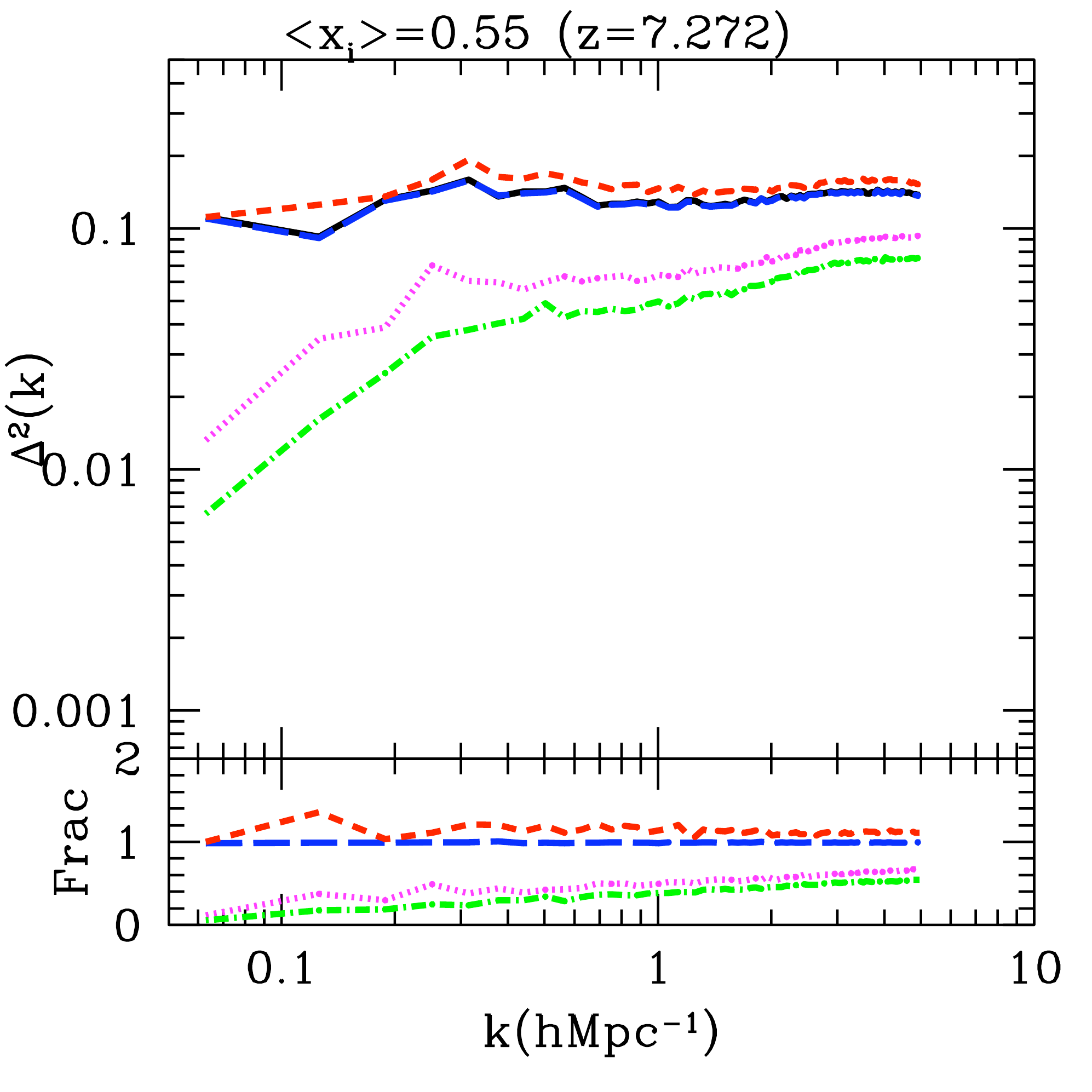}
\end{center}
\vspace{-3mm}
\caption{The predicted 21-cm dimensionless power spectra for the models discussed in this paper. Panels are shown for a range of values of $\left<x_{i}\right>$ corresponding to different stages of reionization as shown in Figures~\ref{HIIevolve1} - \ref{HIIevolveII5}. The units of the dimensionless power spectrum are $(28[(1+z)/10]$mK$)^{2}$. In the lower sub-panels we show the ratio of the the other models to the Bow06 model.} \label{PSA}
\end{figure*}

The results of Figures~\ref{HIIevolve1}-\ref{HIIevolveII5} indicate that the 21cm power spectrum will depend on the galaxy formation model assumed. This is shown in Figure~\ref{PSA} which displays power spectra for each of the semi-analytic models in Table~\ref{Parameters}. From this figure we see that Bow06 and NOSN models show a large variation in 21-cm power spectrum predictions, with the amplitude of 21cm power spectrum for the Bow06 model being higher than the NOSN model across all  wave-numbers (with the exception of very late in the reionization process).  
Secondly, the Bow06 and Lagos models show slightly different 21-cm power spectrum predictions. This is  because the modified  star formation law in the Lagos model relative to the Bow06 model leads to different predictions for the number of luminous galaxies (Figure~\ref{UVLF}) and hence the clustering of the the ionising source population. Thirdly, the NOSN model has a larger amplitude for the 21cm power spectrum than does the NOSN(no suppression) model. This shows that the photo-ionization effect on the 21cm power spectrum can be 
seen in the no SNe feedback models. Conversely, we find negligible difference between the Bow06 and Bow06(no suppression) models. This very small difference means that photo-ionization feedback can only effect the reionization signature in the absence of SNe feedback. These findings represent the main results of the paper.

Figure~\ref{PSA} also shows the evolution of predicted 21-cm dimensionless power spectra for the different models. { The evolution of the power spectrum in Bow06, Bow06(no suppression) and Lagos models show the characteristic rise and fall described in detail by \citet[][]{lidz2008}. The maximum amplitude of the 21-cm power spectrum occurs at a scale of around $k\sim0.2h^{-1}$Mpc for an ionization fraction of $\left<x_{i}\right>\sim0.75$.} In all models there is a trend for the wavenumber $k$ at which the shoulder due HII regions occurs to decrease (corresponding to increasing size of HII regions) with increasing ionisation fraction. The largest  difference is seen between the Bow06 and NOSN models, and is most pronounced at large scales (i.e. small wave numbers). In this regime the NOSN power spectrum is lower than for Bow06. At smaller wavenumbers (i.e. large scales) the Bow06 model has higher amplitude than the NOSN model for all ionization fraction ranges. The difference in amplitude between the two models increases from $\left<x_{i}\right> \sim0.056$ to $\left<x_{i}\right> \sim0.16$, before decreasing later in the reionization history. We cannot distinguish the difference between the Bow06 and the Bow06(no suppression) model at any redshifts. There is a small difference between the Bow06 and Lagos models at all redshifts, although the magnitude of difference is smaller than between the Bow06 and NOSN models. There is also a difference between the NOSN and NOSN(no suppression) models.
The bottom panels in Figure~\ref{PSA} show the ratio between the models and the Bow06 model for each redshift.
 
\subsection{Observational implications}

\begin{figure}
\begin{center}
\includegraphics[width=8cm]{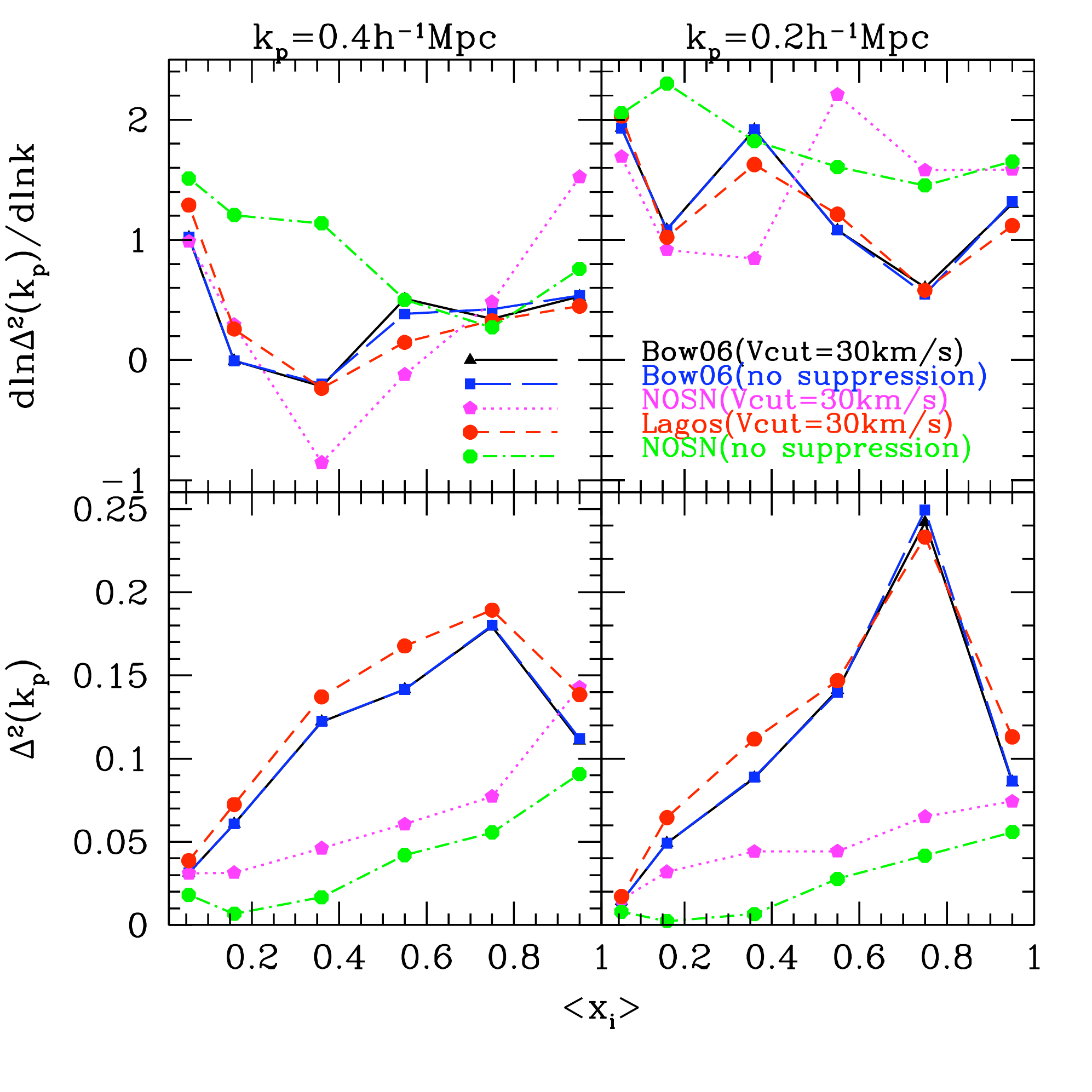}
\end{center}
\vspace{-8mm}
\caption{Plots of the evolution in dimensionless 21cm power spectrum amplitude (lower panels) and slope (upper panels) as a function of ionization fraction $\left<x_{i}\right>$. Predictions are shown for models, Bow06 (triangles, black solid line), NOSN (pentagons, violet dotted line), Bow06(no suppression) (squares, blue long dashed line), Lagos (circles, red dashed line) and NOSN(no suppression) (octagons, green dot dashed line). Results are shown for two central wave numbers, $k_{\rm p}=0.4h^{-1}$Mpc (left) and $0.2h^{-1}$Mpc (right){ , corresponding to the point on the power spectrum at which we evaluate the amplitude and gradient.} } \label{COA}
\end{figure}

\citet[][]{lidz2008} demonstrated that first generation low frequency arrays like the MWA\footnote{www.mwa.org} should have sufficient sensitivity to measure the amplitude and slope of the 21 cm power spectrum. To quantify the effect of SNe feedback on the power spectrum we therefore compare the amplitude and slope of predicted 21-cm power spectra for the Bow06, NOSN, Bow06(no suppression), NOSN(no suppression) and Lagos models. In Figure~\ref{COA},  we plot these values as a function of the $\left<x_{i}\right>$ for central wave numbers of $k_{\rm p}=0.2h^{-1}$Mpc and $0.4h^{-1}$Mpc, corresponding to the range of wave numbers to be probed by the MWA. There are significant differences in the predicted quantities. For $k_{\rm p}=0.2h^{-1}$Mpc, the inclusion of SNe feedback results in fractional changes that are of order unity, particularly near the peak of reionization. 

\begin{figure}
\begin{center}
\includegraphics[width=8cm]{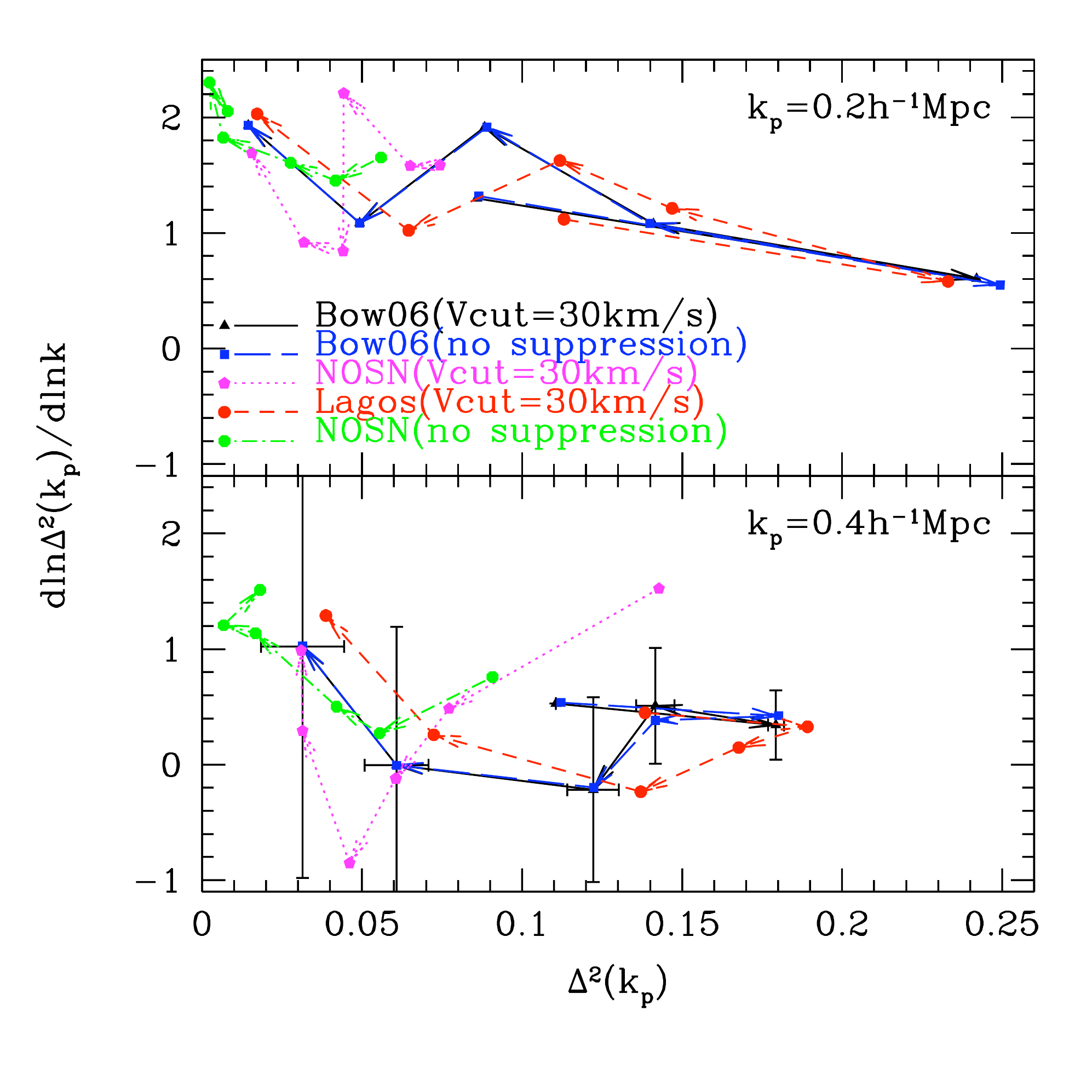}
\end{center}
\vspace{-8mm}
\caption{Plots of the loci of points in the parameter space of 21-cm power spectrum amplitude and slope. Loci are shown for each of 
Bow06 (triangles, black solid line), NOSN (pentagons, violet dotted line), Bow06(no suppression) (squares, blue long dashed line), Lagos (circles, red dashed line) and NOSN(no suppression) (octagons, green dot dashed line) models. %the Bow06 (triangles), NOSN (pentagons), Bow06($V_{\rm cut}$=0km/s) (square) and Lagos(circles) models. 
Results are shown for two central wave numbers, $k_{\rm p}=0.2h^{-1}$Mpc (top) and $0.4h^{-1}$Mpc (bottom){, corresponding to the point on the power spectrum at which we evaluate the amplitude and gradient.}
The error bars for $0.4h^{-1}$Mpc at each point on the Bow06 model correspond to estimates for the MWA ({ specifically an $r^{-2}$ distribution of 500 antennas}) \citep[][]{lidz2008} with 1000 hours of integration and 6MHz of bandpass. }
\label{SLPS}
\end{figure}

Since the ionization fraction is not a direct observable, we plot the progression of a model in the observable plane of power spectrum  amplitude vs slope. These are shown for the variant models in Figure~\ref{SLPS}, again for the two values of central wavenumber { $k_{\rm p}$, corresponding to the point on the power spectrum at which we evaluate the amplitude and gradient.} The arrows show the direction from high to low $\left<x_{i}\right>$ (from 0.95 to 0.056). The tracks separate according to whether SNe feedback is included or not and difference of star formation prescription in the models (the Bow06 and Lagos models). 
To illustrate the potential for detectability of this difference we also include error bars at each point corresponding to estimates for the MWA ({ specifically an $r^{-2}$ distribution of 500 antennas}) \citep[][]{lidz2008} assuming 1000 hours integration and 6MHz bandpasses for wavenumber $k_{\rm p}$=0.4$h^{-1}{\rm Mpc}$. 
The figure demonstrates that mid-way through reionization (i.e. at the highest amplitude), the difference between the tracks for models with and without SNe feedback could be detected by the MWA, indicating that the strength of SNe feedback during the epoch of reionization could be inferred directly from observations of the 21cm power spectrum.    

\section{Summary and conclusions}\label{Summary}

Over the next decade we are likely to see the first measurements of the power spectrum of redshifted 21cm fluctuations from neutral hydrogen structure during the Epoch of Reionization. One goal of these experiments will be to learn about the properties of the galaxies that drove the reionization process. It is known that the ionisation structure of the IGM, and hence the observed 21cm power spectrum will be sensitive to the astrophysical properties of the reionizing galaxies. With this in mind, \citet{barkana2008} has suggested that analytic models of the power spectrum could be used to determine the astrophysics of the reionizing galaxies, provided they are tuned to provide a sufficiently precise description through comparison with numerical simulations. However, previous analyses of the structure of reionization and the predicted power spectrum have used very simple prescriptions to relate ionizing luminosity to the underlying dark matter distribution. In this paper we have made a first attempt to connect the details of the ionisation structure and 21cm power spectrum with realistic models for galaxy formation by combining the GALFORM galaxy formation model implemented within the Millennium-II dark matter simulation with a semi-numerical scheme to describe the resulting ionization structure. While not a true calculation of radiative transfer, semi-numerical models are known to reproduce the main features of the 21cm power spectrum where reionization is driven by UV ionizing sources. Our model includes a single value for the escape fraction of ionising photons from galaxies, independent of halo mass. In reality the escape fraction may depend on mass in a way that could be degenerate with the effect of SNe feedback.

We find that the details of galaxy formation are reflected in differences in the structure of reionization. As a result, each of the assumed star-formation law, radiative feedback and SNe feedback are found to affect 21cm power-spectrum predictions. Our main result is that the details of SNe feedback are most important in modifying HII region evolution, and hence the slope and amplitude of the 21 cm power spectrum. We find that photo-ionization feedback also affects HII region evolution but only in the absence of SNe feedback. Thus, unless SNe feedback is ineffective in high redshift galaxies, the reionization process is not self regulating as has been argued previously \citep[e.g.][]{iliev2007}. This finding is consistent with the work of \citet[][]{Theuns2011}  who studied the photon budget in the context of the global evolution of reionization. We find that measurements of the amplitude and slope of the 21cm power spectrum would be sufficient to determine the level at which SNe feedback operated in high redshift galaxies. 

In this work we have concentrated on the effects of SNe and radiative feedback which are relevant to the galaxies thought to dominate reionization and are accessible to semi-analytic models implemented within the Millennium-II simulation, and we have restricted ourselves to the assumption that the escape fraction is not mass dependent. This study illustrates the important role that semi-analytic models can play in realistic simulation of the connection between ionization structure and the properties of the galactic sources responsible for reionization.  Our paper is the first step in a program to determine how redshifted 21cm observations can be used to probe astrophysics of reionization.

\vspace{5mm}

{\bf Acknowledgments} HSK is supported by a Super-Science Fellowship from the Australian Research Council. The Centre for All-sky
Astrophysics is an Australian Research Council Centre of Excellence, funded by grant CE11E0090.  This work was supported in part by the Science and Technology Facilities
Council rolling grant to the ICC. The Millennium-II Simulation was
carried out by the Virgo Consortium at the supercomputer centre of the
Max Planck Society in Garching.
Calculations for this paper were partly performed on the ICC Cosmology Machine, which is part of the DiRAC Facility jointly funded by STFC, the Large Facilities Capital Fund of BIS,
and Durham University.  

\newcommand{\noopsort}[1]{}

\bibliographystyle{mn2e}

\bibliography{21PII-Re}

\begin{appendix}

\section{Modelling spatially dependent reionization feedback in GALFORM}

This appendix describes how spatially dependent reionization feedback is implemented in GALFORM for the calculations in this paper.
The steps in the modelling are as follows:\\ 

\noindent Step (1) We first run GALFORM at high redshift (from redshift$\sim$20) to find the first resolved HII region using the scheme described in \S~\ref{cells}. We assume $(1+F_{\rm c})/f_{\rm esc}=1$ for this step.  \\

\noindent Step (2) For snapshots at redshifts where the first HII region is identified, we then find those galaxies which are inside HII regions and subject to radiative feedback through regulation of cooling processes (i.e. galaxies with $V<V_{\rm cut}$).\\

\noindent Step (3)  A table listing these galaxies is generated for each snapshot. 
%In constructing this table, we use both the value of $ident_{\rm final}$, which is unique for galaxies in the same halo merger tree (rooted at z=0), and $jm$ which is unique for galaxies residing in a tree generated in the $jm^{th}$ mass interval from the GALFORM output. Together the two values $ident_{\rm final}$ and $jm$ identify 
This table identifies which galaxies should have star formation regulated by radiative feedback during subsequent evolution up to the next
snapshot redshift. \\

\noindent Step (4) GALFORM is then run to the next redshift snapshot, including regulation of cooling processes in the galaxies identified in the table at step (3).\\
 
\noindent Step (5) Steps (2) to (4) are then repeated for all snapshots down to redshift 6 where reionization is finished.\\

\end{appendix}

\label{lastpage}
\end{document}